\newcommand{\rem}[1]{}
\newcommand{\eqref}[1]{(\ref{#1})}
\documentclass[12pt]{iopart}

\usepackage{graphicx} % standard LaTeX graphics ti include figure files
\usepackage{cite}     % used for sorting reference numbers in \cite
\usepackage{ulem}
\usepackage{color}

\begin{document}

\title{Friction and Nonlinear Dynamics}

\author{
  N. Manini$^{\rm a}$, O.M. Braun$^{\rm b}$, E. Tosatti$^{\rm c,d,e}$,
  R. Guerra$^{\rm c,d}$, and A. Vanossi$^{\rm d,c}$
}

\address{$^{a}$Dipartimento di Fisica, Universit\`a degli Studi di
  Milano, Via Celoria 16, 20133 Milano, Italy
  }
\address{
$^{b}$Institute of Physics, National Academy of Sciences of Ukraine, 46
  Science Avenue, Kiev 03028, Ukraine
}
\address{
$^{c}$International School for Advanced Studies (SISSA), Via Bonomea
  265, 34136 Trieste, Italy
}
\address{
$^{d}$CNR-IOM Democritos National Simulation Center, Via Bonomea 265,
  34136 Trieste, Italy
}
\address{
$^{e}$International Center for Theoretical Physics (ICTP), Strada
  Costiera 11, 34151 Trieste, Italy
  }

\begin{abstract}
  The nonlinear dynamics associated with sliding friction forms a broad
  interdisciplinary research field that involves complex dynamical
  processes and patterns covering a broad range of time and length scales.
  Progress in experimental techniques and computational resources has
  stimulated the development of more refined and accurate mathematical and
  numerical models, capable of capturing many of the essentially nonlinear
  phenomena involved in friction.
\end{abstract}


%\begin{keywords}nanotribology; sliding friction; nonlinear dynamics;
%  earthquake-like models; Frenkel-Kontorova model; molecular dynamics
%\end{keywords}

\maketitle

%==================================================================
\section{Introduction}
\label{introduction:sec}

Friction phenomena take place across a broad range of time and length
scales, from microscopic atomistic processes, as in the gliding motion of a
nanocluster or a nanomotor \cite{Urbakh04}, up to extremely macroscopic
instances, as in fault dynamics and earthquake events \cite{Scholz98}.
Due to the ubiquitous nature of mechanical dissipative processes and the
enormous practical relevance, friction has been investigated over the
centuries.
While the empirical laws of macroscopic friction are well known
\cite{Bowden50}, the fundamental understanding of the tribological
phenomena at the microscopic scales is still lacking from many points of
view.
The basic difficulty of friction is intrinsic, involving the dissipative
dynamics of large systems, often across ill-characterized interfaces,
and generally violent and nonlinear.
The severity of the task is also related to the experimental difficulty to
probe systems with many degrees of freedom under a forced spatial
confinement, that leaves very limited access to probing the buried sliding
interface.
Thanks to remarkable developments in nanotechnology, new inroads are being
pursued and new discoveries are being made.
At the nanometer scale, state-of-the-art ultra-high-vacuum systems and
local probe studies show a dynamical behavior which is often significantly
different, not just quantitatively but even qualitatively, from the ones
observed in macroscopic tribology.
Bridging the gap among the different length scales in tribological
systems still remains an open challenge.
The phenomenological descriptions that apply to macroscopic friction cannot
yet be derived from the fundamental atomic principles and the interplay of
processes occurring at the molecular level.
Nanofriction is in somewhat better shape.
Together with the current experimental possibility to perform well-defined
measurements on well-characterized materials at the fundamental microscopic
level of investigation of the sliding contacts, advances in the computer
modeling of interatomic interactions in materials science and complex
systems encompass molecular-dynamics (MD) simulations of medium to large
scale for the exploration of the tribo-dynamics with atomic resolution
\cite{PerssonBook, VanossiRMP13}.
Despite the benefits brought about by numerical simulations of realistic 3D
sliding systems, the resulting proliferation of detailed complex data, and
the requirement of always-growing computational efforts have stimulated, in
parallel, the concurrent search for simpler modeling schemes, such as,
e.g., generalized Prandtl-Tomlinson (PT), Frenkel-Kontorova (FK), for
nanofriction, and of Burridge-Knopoff and earthquake-like models, for
mesoscale and macroscale friction, suitable to describe the essence of the
physics involved in highly nonlinear and non-equilibrium tribological
phenomena in a more immediate fashion.

Here we discuss current progress and open problems in the simulation and
modeling of tribology at the microscopic scale, and its connection to the
macroscale.
Neither the PT model, described in detail in several surveys
\cite{Muser03,VanossiRMP13} with several applications to concrete
tip-based physical systems, nor the phenomenological approach based on
the rate-and-state models \cite{Marone98} will be considered here.
With a view to emphasize the role of nonlinearity the present topical
review will restrict to the following theoretical approaches to sliding
friction.
Section~\ref{linear:sec} revises the simple case of near-equilibrium linear
friction in classical mechanics.
Section~\ref{FK:sec} focuses on nonlinearity in crystal sliding in the
framework of the FK model and its generalizations.
Atomistic models and MD nanofriction simulations are presented in
Sec.~\ref{MD:sec}.
Mesoscopic multicontact earthquake-like models are finally examined in
Sec.~\ref{multicontact:sec}.

%=============================================================================
\section{Linear Friction and Dissipation}
\label{linear:sec}

Statistical mechanics accounts for the intimate mechanism of friction: a
system at equilibrium has its kinetic energy uniformly distributed among
all its degrees of freedom.
A sliding macroscopic object clearly is not at equilibrium: one of its
degrees of freedom (the center-of-mass motion) has far more kinetic energy
than any other.
The tendency of the system toward equilibrium will lead to the transfer of
energy from that degree of freedom to all other ones: as a result the
macroscopic object will slow down and its energy will be transferred to the
disordered motion of the other degrees of freedom, resulting in warming up.
This is all sliding friction really is: the tendency of systems toward
equilibrium energy equipartitioning among many interacting degrees of
freedom.
%
% A moving object slows down through collisions with atoms and molecules in
% its surroundings: the object’s kinetic energy is redistributed among a huge
% number of microscopic ``internal'' degrees of freedom forming a ``heat
% bath''.

Thus, in the course of friction under an applied external force, energy is
reversed into the system in the form of frictional heat.
The frictional heat is generally dissipated by some form of heat bath, such
as that provided by a thermostat at temperature $T$.
In a frictional steady state, caused for example by submitting a slider to
an external force $F$, the slider dissipates energy to the bath, and
therefore does not accelerate indefinitely --- it reaches instead a steady
state characterized by an average drift velocity $\langle v \rangle$.
When both $\langle v \rangle$ and $F$ are infinitesimal, the relationship
between the two quantities is linear,
\begin{equation}
\label{Erio-1}
F = m \gamma \langle v \rangle
\,.
\end{equation}
In this so-called ``viscous friction'' the proportionality constant
$\gamma$ is the linear friction coefficient.
It is known from classical statistical mechanics, for example of Brownian
motion as described by the Langevin equation, that for linear friction
systems which obey Eq.~(\ref{Erio-1}), the Einstein relationship
\begin{equation}
\label{Erio-2}
D \gamma = k_B T /m
\end{equation}
is generally valid, connecting the friction coefficient $\gamma$, which
measures dissipation, to the diffusion coefficient $D$, which measures
fluctuations.
This expresses the fluctuation-dissipation theorem of linear, viscous
friction.

%-----------------------------------------------------------------------------
\subsection{The Fluctuation-Dissipation Theorem}\label{fluctdiss:sec}

To simulate the classical motion of a macroscopic object moving in
contact with an equilibrium bath such as the molecules of a gas or a
liquid, or the phonons of a solid, the standard implementation
\cite{Gardiner,Frenkel-Smit96} requires adapting Newton's equations of
motion with the addition of a damping force $\vec{f}_{\rm damp}$ plus a
random force $\vec{f}_{\rm rand}(t)$.
The damping force represents the transfer of energy from the macroscopic
object of mass $m$ to the heat bath, i.e.\ dissipation:
\begin{equation}\label{eq_Langevin}
  \vec{f}_{\rm damp} =  - m \gamma \dot{\vec{r}} \,.
\end{equation}
This formula, equivalent to Eq.~\eqref{Erio-1}, assumes that the deviation
from equilibrium is small, so that {\em linear response} holds: the
restoring Stokes force is linear in the perturbing velocity, and acts
opposite to it to restore the $\langle\dot{\vec{r}}\rangle =0$ equilibrium
regime.
This linear dependence is purely the lowest order term in a Taylor
expansion: there is no reason to expect the linear relation
\eqref{eq_Langevin} to extend to large velocity, and indeed e.g.\ the drag
friction of speeding objects in gases is well known to follow Rayleigh's
quadratic dependence on speed $|\dot{\vec{r}}|$.
The random-force term represents statistically the ``kicks'' that the
objects experiences due to its interaction with the thermal bath.
In the frame of reference of the thermal bath
$\langle\vec{f}_{\rm rand}(t)\rangle =\vec 0$  of course.
The random term is the result of many very frequent collisions events,
resulting in random forces uncorrelated with themselves except over very
short time spans.
More precisely, we assume there is some maximum time $\tau$ beyond which
any correlation vanish:
\begin{equation}\label{bathuncorr}
  \langle f_{{\rm rand}\,\alpha}(t)f_{{\rm rand}\,\alpha'}(t+\delta t)\rangle
  =0 \quad {\rm if}\ \delta t>\tau
  \,.
\end{equation}
In addition, the assumption of thermal equilibrium ensures us that the bath
is in a steady state, so that $\langle f_{{\rm rand}\,\alpha}(t)f_{{\rm
    rand}\,\alpha'}(t+\delta t)\rangle$ is independent of $t$, and depends
on $\delta t$ only: the statistical properties of the random force are
constant in time.
In most practical situations, one can safely ignore the dynamics over a
time scale of the order of $\tau$ or shorter.
We are interested instead in the integral effect of $\vec{f}_{\rm
  rand}(t)$ over some time period $t$ that is long compared to $\tau$.
We can break up that integral into many pieces, each covering a duration
$\tau$:
\begin{equation}
  \int_0^t f_{{\rm rand}\,\alpha}(t')dt'=
  \int_0^\tau f_{{\rm rand}\,\alpha}(t')dt'+
  \int_\tau^{2\tau} f_{{\rm rand}\,\alpha}(t')dt'+
  \int_{2\tau}^{3\tau} f_{{\rm rand}\,\alpha}(t')dt'+\dots
\end{equation}
This integral is then a sum of many independent random terms, each drawn
from the same distribution whose only relevant property is that it has zero
mean value.
As a result of the central-limit theorem, the total integral obeys a
Gaussian distribution with null mean, and whose standard deviation scales
with the number of terms in the sum, i.e.\ $t^{1/2}$.

By taking the equation of motion in the absence of any external driving,
\begin{equation}\label{eq_diffusion}
  m\ddot{\vec{r}} = \vec{f}_{\rm damp} + \vec{f}_{\rm rand}(t)
  \,,
\end{equation}
and integrating it in time we obtain
\begin{equation}\label{eq_diffint}
  \dot{\vec{r}}(t) = \dot{\vec{r}}(0) \, e^{-\gamma t}
  + \frac 1m \int_0^t e^{-\gamma (t-t')}\, \vec{f}_{\rm rand}(t')\, dt'
  \,.
\end{equation}
The first term at the right-hand side of Eq.~\eqref{eq_diffint} becomes
negligible for a time $t\gg 1/\gamma$, long enough for the object to
equilibrate with the thermostat and lose memory of its initial
condition.
In this large-$t$ limit, by taking the square module of
Eq.~\eqref{eq_diffint} and executing the ensemble average, we have
\begin{equation}\label{dotsq}
  \lim_{t\to\infty}\langle \dot{\vec{r}}^{\,2}(t)\rangle =
  \frac 1{m^2} \lim_{t\to\infty}
  \int_0^t e^{-\gamma (t-t')}\,
  \int_0^t e^{-\gamma (t-t'')}\,
  \langle\vec{f}_{\rm rand}(t') \cdot
  \vec{f}_{\rm rand}(t'')\rangle
  \, dt'dt''
  \,.
\end{equation}
The term at the left side multiplied by ${1\over 2} m$ %/2$
yields the average kinetic
energy, which by standard equipartition needs to equal $3k_{\rm B} T/2$.
By rearranging the exponentials on the right-hand side and substituting
$t_1=t-t''$ and $t_2=t''-t'$, we obtain:
\begin{eqnarray}\nonumber
  \frac 32 k_{\rm B} T &=&
  \frac 1{2m} \lim_{t\to\infty}
  \int_0^t \int_0^t e^{-\gamma (2t-t''-t')}\,
  \langle\vec{f}_{\rm rand}(t')\cdot
  \vec{f}_{\rm rand}(t'')\rangle
  \, dt'dt''\\\nonumber
  &=&
  \frac 1{2m} \lim_{t\to\infty}
  \int_0^t \int_0^t e^{-\gamma (2t-2t''+t''-t')}\,
  \langle\vec{f}_{\rm rand}(0)\cdot
  \vec{f}_{\rm rand}(t''-t')\rangle
  \, dt'dt''
  \\\nonumber
  &=&
  \frac 1{2m} \lim_{t\to\infty}
  \int_0^t dt_1 \int_{-t}^t dt_2 \, e^{-\gamma (2t_1+t_2)}\,
  \langle\vec{f}_{\rm rand}(0)\cdot
  \vec{f}_{\rm rand}(t_2)\rangle
  \\\nonumber
  &=&
  \frac 1{2m}
  \int_0^\infty dt_1 e^{-\gamma 2t_1}\int_{-\infty}^\infty dt_2 \,e^{-\gamma t_2}\,
  \langle\vec{f}_{\rm rand}(0)\cdot
  \vec{f}_{\rm rand}(t_2)\rangle
  \\\label{equipart}
  &=&
  \frac 1{4m\gamma}
  \int_{-\infty}^\infty dt_2 \,e^{-\gamma t_2}\,
  \langle\vec{f}_{\rm rand}(0)\cdot
  \vec{f}_{\rm rand}(t_2)\rangle
  \,,
\end{eqnarray}
where we have used the steadiness of the stochastic process discussed
after Eq.~\eqref{bathuncorr}.
Assuming, as is commonly the case, that the autocorrelation time $\tau$
of the random term is short compared to $\gamma^{-1}$, the integrand of
Eq.~\eqref{equipart} has $e^{-\gamma t_2}\simeq 1$ in all region
$|t_2|\leq \tau$ of delays where the factor $\langle\vec{f}_{\rm
  rand}(0)\cdot \vec{f}_{\rm rand}(t_2)\rangle$ is significantly
different from $0$.
This observation further simplifies Eq.~\eqref{equipart} to the {\em
  fluctuation-dissipation} relation
\begin{equation}\label{fluctdiss}
  6 m \gamma k_{\rm B} T =
  \int_{-\infty}^\infty dt_2 \,
  \langle\vec{f}_{\rm rand}(0)\cdot
  \vec{f}_{\rm rand}(t_2)\rangle
  \,.
\end{equation}
This expression draws an explicit link between the autocorrelation
amplitude of the fluctuations and the product of the dissipation
coefficient and thermostat temperature.
Note that the integral in Eq.~\eqref{fluctdiss} depends on both the
amplitude of fluctuations $\vec{f}_{\rm rand}$ and the time over which
they remain self-correlated.
The effect on the mesoscale object increases if the random force is
larger and/or if the time interval over which $\vec{f}_{\rm rand}$
pushes in the same direction before changing is longer.
The relation \eqref{fluctdiss} can be equally well satisfied by weaker
random forces acting for longer correlation times or stronger forces with
shorter $\tau$, which thus lead to the same statistical effects.
As $\tau$ is the shortest time scale around, for all practical purpose
one can satisfy Eq.~\eqref{fluctdiss} assuming a sort of $\tau\to 0$
limit:
\begin{equation}\label{eq_Gaussian}
  \langle f_{{\rm rand}\,\alpha}(t)f_{{\rm rand}\,\alpha'}(t')\rangle =
  2m\gamma k_{\rm B}T\delta_{\alpha\alpha'}\delta(t-t')
  \,,
\end{equation}
providing a simple recipe for computer simulations.
For simulations of models such as the PT or FK ones, the
phenomenological degrees of freedom are often coupled to a Langevin
thermostat of this kind, implying that each degree of freedom is
actually coupled to a vast number of other bath degrees of freedom.
Even in MD simulations of atomic-scale friction, Langevin thermostats
are applied to all or a part of the atoms involved \cite{Frenkel-Smit96,
  Robbins01, Muser06}.
Of course, this approach is not rigorous, since the relevant particles
colliding with each given simulated atom are already all included in the
conservative and deterministic forces explicitly accounted for by the
``force field''.
The Langevin approach is quite accurate to describe small perturbations
away from equilibrium, but it may fail quite badly in the strongly
out-of-equilibrium nonlinear phenomena which are the target of the present
paper.

%-----------------------------------------------------------------------------
\subsection{Linear versus Nonlinear Friction}

In the rest of this review we will deal with nonlinear frictional
phenomena, which deviate violently from linearity and near-equilibrium,
and where therefore Eqs.~(\ref{Erio-1}) and~(\ref{Erio-2}) do not
generally apply.
As it has, surprisingly, only been realized in the last few decades, even
arbitrarily violently non-equilibrium and nonlinear driven phenomena adhere
to an extension of the fluctuation-dissipation theorem.
That is the Jarzynski (or Jarzynski--Crooks) relation
\cite{Crooks98,Jarzynski07}, whose simplest form can be briefly summarized
as follows.

Suppose starting from a system in state A at temperature $T$, and apply an
external force of arbitrary form and strength causing it to evolve, for
example to slide, to another state B; assume for simplicity B to be also a
state of equilibrium.
Call $W_{AB}$ the work done by the external force, and call $\Delta {\cal
  F}_{AB} = {\cal F}_B - {\cal F}_A$ the difference of equilibrium free
energy between the states B and A.
Clearly, $\langle W_{AB} \rangle > \Delta {\cal F}_{AB}$ must be valid, because some
work will be wasted in going from A to B, unless that was done
infinitely slowly (adiabatically).

Suppose now repeating the forced motion A$\to$B many times.
Each time,  $W_{AB}$ will be different.
The Jarzynski equality states that
\begin{equation}
\label{Erio-4}
\langle \exp (- W_{AB} / k_B T) \rangle = \exp (- \Delta {\cal F}_{AB} /k_B T)
\,.
\end{equation}
It can be shown that in near-equilibrium conditions, Eq.~(\ref{Erio-4})
is completely equivalent to the fluctuation-dissipation theorem.
The beauty of it is however that Eq.~(\ref{Erio-4}) is totally general.

One particular case is useful in order to underline its far-reaching
power.
Suppose to take B=A, that is a final state identical to the starting one.
In that case
\begin{equation}
\label{Erio-5}
\langle \exp (- W_{AA} / k_B T) \rangle =1
\,.
\end{equation}
This equation appears at first sight impossible to satisfy, because surely
all $\langle W_{AA} \rangle > 0$: all forced motion must cost work.
The answer is that the probability $P(W_{AA})$ is indeed a distribution
centered around a positive $W_{AA} > 0$, but with a nonzero tail
extending to $W_{AA} < 0$.
This tail represents rare events where work is gained rather than spent
--- we can think of them as a sort of ``free lunches''.
% They are a strong and tangible manifestation of the role of
% fluctuations in arbitrarily nonlinear,
% dissipative and frictional phenomena.
Jarzynski's theorem requires that ``free lunches'' must occur precisely
in such a measure to satisfy Eq.~(\ref{Erio-5}).
However it is easy to convince ourselves that they will be frequent only
in microscopic systems, where $P(W_{AA})$ is broad.
The larger the system involved, the narrower $P(W_{AA})$ will be, the
rarer and rarer the occurrence of ``free lunches''.
In a macroscopic friction experiment, the occurrence of a ``free lunch''
will be virtually impossible.

%=============================================================================
\section{The Frenkel-Kontorova Model}
\label{FK:sec}

In nanoscale tribology, extensive attention has focused on the time-honored
PT model, which describes a point-like tip sliding over a space-periodic
crystalline surface in a minimal fashion.
We shall omit this model from the present review, since it is covered in
great detail elsewhere \cite{Muser03,VanossiRMP13,Popov14}.
We concentrate instead on its natural extension, the one-dimensional FK
model \cite{Braunbook}, which provides a prototypical description of the
mutual sliding of two perfect, extended crystalline surfaces.
First studied analytically in Ref.~\cite{Dehlinger29} and later introduced
independently to address the dynamics of dislocations in crystals
\cite{Frenkel38, Kontorova38a, Kontorova38b}, subsequently this model
became the paradigm describing the structure and dynamics of adsorbed
monolayers in the context of surface physics.

\begin{figure}
\centerline{
\includegraphics[width=0.5\textwidth,clip=]{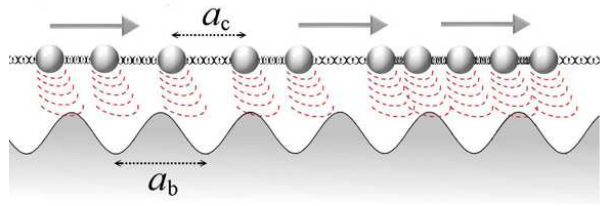}
}
\caption{\label{fig:FKmodel}
 A sketch of the FK model, showing the two competing lengths: the average
 interparticle spacing and the lattice periodicity of the substrate.
}
\end{figure}

The standard FK model consists of a 1D chain of $N$ classical particles
(``atoms''), interacting via harmonic forces and moving in a sinusoidal
potential, as sketched in Fig.~\ref{fig:FKmodel}.
The Hamiltonian is
\begin{eqnarray}\label{FKHamil}
H = \sum_{i=1}^N
\left[ \frac{{{p_i}^2}}{{2m}} + \frac{1}{2} K (x_{i + 1}  - x_i  - a_c)^2
+ \frac{1}{2} U_0 \cos \frac{2 \pi x_i}{a_b}  \right] \,.
\end{eqnarray}
In Eq.~\eqref{FKHamil}, the $p_i^2/(2m)$ term represents the kinetic energy
of the particles, and the next term describes the harmonic interaction,
with elastic constant $K$, of nearest-neighboring atoms at equilibrium
distance $a_c$.
The final cosine term describes the ``substrate corrugation'', i.e.\ the
periodic potential of amplitude $U_0$ and period $a_b$, as experienced
by all particles alike.
To probe static friction, all atoms are driven by an external force $F$,
which is increased adiabatically until sliding starts.

The continuum limit of the FK model, appropriate for large $K$, is the
exactly integrable sine-Gordon (SG) equation, and this mapping contributed
to the great success of the FK model.
The solutions of the SG model include nonlinear topological solitons (known
as ``kinks'' and ``antikinks''), plus dynamical solitons (``breathers''),
beside linear vibration waves (phonons).
In the FK model, the sliding processes are entirely governed by its
topological excitations, the kinks.
Let us consider the simplest ``commensurate'' case, where before sliding
the chain is in a trivial ground state (GS), when $N$ atoms fit one in each
of the $M$ minima of the substrate potential, so that the coverage
(i.e.\ the relative atomic concentration) $\theta =N/M=a_b/a_c$ equals $1$.
In this case, the addition (or subtraction) of a single atom results in
configurations of the chain characterized by one kink (or antikink)
excitation.
Still at zero applied force, in order to reach a local minimum of the total
potential energy in Eq.~\eqref{FKHamil}, the kink expands in space over a
finite length, so that the resulting relaxed chain configuration consists
in a local compression (or expansion, for an antikink).
Upon application of a force, it is far easier to move along the chain for
kinks than for atoms, since the activation energy $\varepsilon_{\rm PN}$
for a kink displacement [known as the Peierls-Nabarro (PN) barrier] is
systematically smaller, and often much smaller \cite{Floria96}, than the
amplitude $U_0$ of the energy barrier that single atoms experience in the
substrate corrugation.

\begin{figure}
\centerline{
\includegraphics[width=0.6\textwidth,clip=]{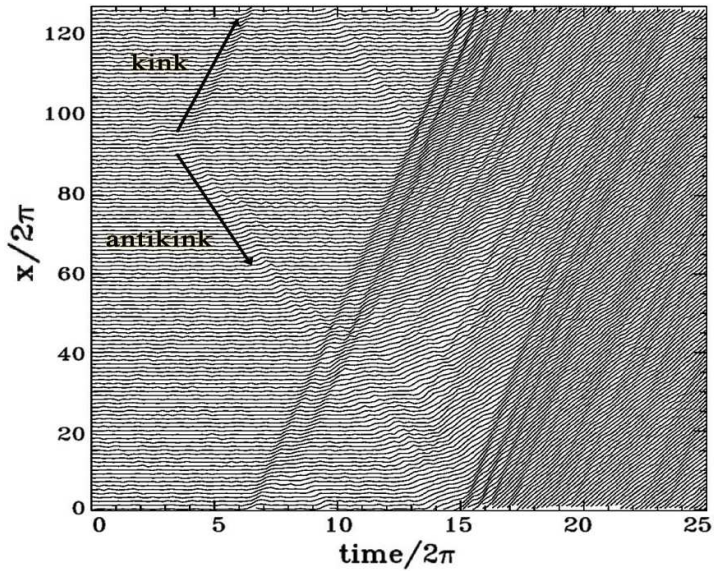}
}
\caption{\label{kinks}
  Time dependence of the atomic trajectories for the fully matched
  ($\theta=1$) FK model at the (low-temperature) onset of depinning.
  Motion starts with the nucleation of a kink-antikink pair.
  The kink and the antikink depart in opposite directions cross the
  periodic boundary conditions, and collide quasielastically.
  A second kink-antikink pair forms in the wake of the initial kink.
  Further kink-antikink pairs are generated, with an avalanche-like
  increase of the kink-antikink concentration, eventually leading to a
  sliding state.
  Adapted from Ref.~\cite{Braun97}, Copyright (1997) by The American
  Physical Society.
}
\end{figure}

The motion of kinks (antikinks), i.e.\ the displacement of the extra atoms
(vacancies) represents the mechanism for mass transport along the chain.
These displacements are responsible for the mobility, diffusivity, and
conductivity within this model.
Generally therefore a larger concentration of kinks is associated to a
larger the overall mobility \cite{Vanossi03}.
For the simple commensurate GS (e.g., $\theta =1$), which contains neither
kinks nor antikinks, the onset of sliding motion under a driving force
occurs via the creation of a kink-antikink pair, e.g.\ induced by a thermal
fluctuation, see Fig.~\ref{kinks}.

If the FK chain is of finite length, kinks/antikinks are usually created at
one chain free end, then they advance along the chain, eventually
disappearing at the opposite end \cite{Braun90b}.
Every kink running from one end of the chain to the other produces the
advancement of the entire chain by one lattice spacing $a_b$.
For a finite film confined between two surfaces, or for an island deposited
on a surface, the general expectation is that sliding initiates likewise
with the formation and entrance of a kink, or antikink at the boundary
\cite{Varini15}.
In this two-dimensional (2D) case, and more generally in $D$-dimensional
systems, the zero-dimensional kinks of the FK model are replaced by $(D-1)$
dimensional misfit dislocations or domain walls, whose qualitative physics
and role is essentially the same.

Incommensurability between the periods $a_b$ and $a_c$ plays an important
role in the FK model.
Assume, in the limit of an infinite chain length, the ratio $\theta
=a_b/a_c$ of the substrate period $a_b$ to the average spacing $a_c$ of the
chain to be irrational.
The GS of the resulting incommensurate FK model is characterized by a
sort of ``staircase'' deformation, with a regular sequence of regions
where the chain is compressed (or expanded) to match the periodic
potential, separated by kinks (or antikinks), where, at regular
intervals, the misfit stress is released through a localized expansion
(compression).
The incommensurate FK model exhibits, under fairly general conditions on
$\theta$ \cite{Meiss92}, a critical elastic constant $K= K_c$, such that if
$K > K_c$ the chain can slide freely on the substrate at no energy cost,
i.e.\ the static friction $F_s$ drops to zero (and the low-velocity kinetic
friction becomes extremely small), while, remarkably, this is no longer
true when $K < K_c$.
In the early 1980s a rigorous mathematical theory of this phenomenon called
``the transition by breaking of analyticity'', now widely known as the {\em
  Aubry transition} \cite{Ying71, Sokoloff77, Pokrovsky78, AubrySolitons,
  AubryLeDaeron, Peyrard83}, was developed.
A simple explanation of free sliding in the unpinned $F_s =0$ state is
the following.
For every atom climbing up toward a corrugation potential maximum, there
always is another atom moving down, with an exact energy balance of these
processes.
Quite generally, incommensurability guarantees that the total energy (we
are at $T=0$) is independent of the relative position of the chain and
the periodic lattice.
However, in order for the chain to slide with continuity between two
consecutive positions, it is necessary that particles should be able to
occupy a maximum of the potential, the worst possible position.
At the Aubry transition, however, realized by a relative increase of the
periodic potential magnitude, or equivalently by a softening of the
chain stiffness, the probability for a particle to occupy that position
drops from a finite value to exactly zero.
The nature of this transition, which is structural but without any other
static order parameter (besides energy, of course), is dynamical, similar
in that to a glass transition: simply, a part of phase space becomes
unavailable, in this case by sliding.
The chain is unpinned and mobile as long as, in its GS, atoms may occupy
with a finite probability all positions, including those arbitrarily
close to the maxima of the substrate potential, but is immobilized when
that possibility ceases.
The critical chain stiffness $K= K_c$ marks the crossing of the Aubry
transition, where the chain turns from the free sliding state to the locked
(``pinned'') state with a nonzero static friction $F_s$ \cite{Floria96}.
The value $K_c$ is in turn a discontinuous function on the length ratio
$\theta$ characterizing the model.
The minimum value $K_c \simeq 1.0291926$ [in units of $2 U_0 (\pi/a_b)^2$]
is achieved for the golden-mean ratio $\theta= (1+ \sqrt{5})/2$
\cite{Braunbook}.
The stiff-spring chain with $K>K_c$ can explore adiabatically the full
infinite and continuous set of GSs configurations by means of displacements
at no energy cost.
This zero-frequency freely-sliding mode is the Goldstone mode consistent
with an emerging continuous translational invariance of the model,
connecting continuously with an acoustical phase mode (phason) at finite
wavelength.
By contrast, in the pinned soft-chain region $K<K_c$, all particles
remain trapped close to the substrate-potential minima, a configuration
which exhibits a finite energy barrier against motion over the
corrugation.

The locking is provided here, despite translational invariance, by the
inaccessibility of forbidden configurations, which act as dynamical
constraints.
Above $K_c$ the incommensurate-chain sliding can therefore be initiated by
an arbitrarily small driving force, whereas for $K < K_c$ the chain and the
corrugation lock together through the pinning of the kinks or superkinks
separating locally lattice-matched regions.
Note that the locking of the free ends in a finite-size FK chain
necessarily leads to pinning, even when $\theta$ is irrational and
regardless of how large $K$ is.
Even for a finite chain, it is nevertheless still possible to define and
detect a symmetry-breaking Aubry-like transition \cite{Braiman90,
  Benassi11a, Pruttivarasin11}.

\begin{figure}
\centerline{
\includegraphics[width=0.6\textwidth,clip=]{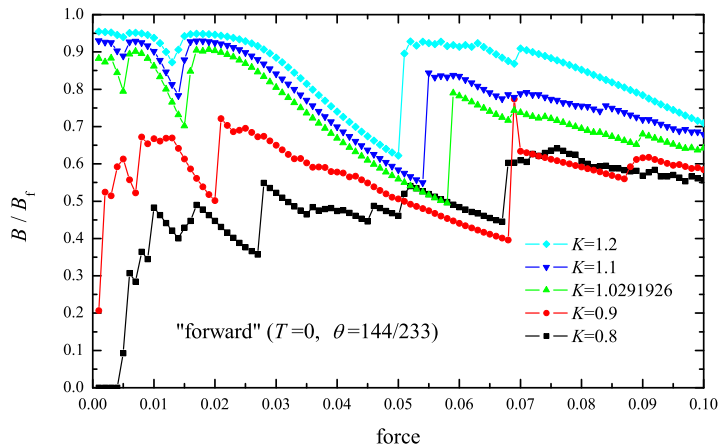}
}
\caption{\label{fig:AubryFig}
  As a function of the dc force $F$, the zero-temperature mobility $B=v/F$
  normalized to the free-motion value $B_f=(m \gamma)^{-1}$ for the
  standard FK model with the golden-mean length ratio for a few values of
  the elastic constant $K$ above and below the Aubry threshold $K_c$.
  The equations of motion include a viscous damping, with a damping % friction
  coefficient $\gamma=0.1$.
  Adapted from Ref.~\cite{Braunbook}.
}
\end{figure}

For the characterization of the Aubry transition, a ``disorder'' parameter
$\psi$ was conveniently defined \cite{Coppersmith83} as the smallest
distance of atoms from the nearest maximum of the corrugation potential.
This quantity vanishes in the freely-sliding state, and is nonzero in the
pinned state.
At the critical pinned-to-sliding point, the disorder parameter exhibits a
power-law behavior
\begin{equation}
\label{criticalexp:eq}
  \psi \propto (K_c - K)^{\chi_{\psi}},
\qquad
  F_s \propto \varepsilon_{\rm PN} \propto (K_c - K)^{\chi_{\rm PN}}.
\end{equation}
Here the values of the critical exponents are functions of the irrational
length ratio $\theta$.
Specifically, for the golden-mean ratio
$\chi_{\psi} \simeq 0.7120835$ and
$\chi _{\rm PN} \simeq 3.0117222$ \cite{Peyrard83, LinHu92, MacKay93,
  Coppersmith83, Sharma84, deSeze84, Aubry89, MacKay91}.
Equation~\eqref{criticalexp:eq} characterizes the continuous Aubry
transition with a scaling behavior typical of critical phenomena, here at
$T=0$ but as a function of the the stiffness parameter $K$.
It is common to refer to the exponents in Eq.~\eqref{criticalexp:eq} as
super-critical, since they are specific to the pinned side of the
transition, $K\leq K_c$.

Sub-critical exponents were introduced for the freely-sliding state
$K>K_c$, as well.
To describe the response of the model to an infinitesimally small dc force
$F$ applied to all atoms, an extra damping term $-m \gamma \dot x_i$ has to
be included in the equation of motion to prevent unlimited acceleration, and
to achieve instead a steady-state.
The resulting effective viscosity in the subcritical region is defined as
$\Gamma = \lim_{F \rightarrow 0} F/(mv)$, in terms of steady-state average
velocity $v$ resulting in response to $F$.
At the Aubry critical point $K_c$, the effective viscosity $\Gamma$
diverges.
For $\theta=(1+5^{1/2})/2$, the golden-mean ratio, the scaling behavior of
$\Gamma$ is
\begin{equation}
\label{GammaAubry}
\Gamma (K) \propto (K-K_c)^{-\chi_{\Gamma}}
\,,
\end{equation}
with $\chi_{\Gamma} \simeq 0.029500$.
As is the case for all scaling relations, Eq.~(\ref{GammaAubry}) provides
the leading divergence close to the Aubry point; at a larger distance from
$K_c$, $\Gamma$ deviates from Eq.~(\ref{GammaAubry}).
Eventually, in the $K\to\infty$ SG limit, $\Gamma$ decreases toward
$\gamma$.
In general, in the unpinned phase at $K>K_c$ the incommensurate FK model
exhibits an effective viscosity systematically larger and thus a mobility
$B=v/F=(m\Gamma)^{-1}$ consistently smaller than its maximum value
$(m\gamma)^{-1}$.
This observation is illustrated by the $F \to 0$ limiting values of the
$K>K_c$ curves of Fig.~\ref{fig:AubryFig}.
Exclusively in the $K \to \infty$ SG limit, the incommensurate system moves
under an infinitesimal force without any extra dissipation added to the
base value $\gamma$, therefore in a frictionless sliding motion, despite
the finite corrugation magnitude $U_0$.
The first prediction of vanishing static friction was formulated for the
incommensurate infinite-size sufficiently hard FK chain by Peyrard and
Aubry \cite{Peyrard83}.
This phenomenon was subsequently re-discovered for incommensurate
tribo-contacts, and named {\it superlubricity} \cite{Hirano90, Shinjo93}.
This name has drawn criticism, because it could misleadingly suggest the
vanishing of {\em kinetic} friction too, in analogy to superfluidity or
superconductivity.
Actually instead, the depinning of sliding interfaces closes just one of
the channels for energy dissipation, namely the one associated to the
stick-slip elastic instability at low speed.
Additional dissipation channels, including the emission of vibrations such
as sound waves, remain active, with the result that the actual kinetic
friction force remains nonzero and growing with increasing sliding speed.
All the same, the superlubric state does attain a significant reduction of
the kinetic friction force (and thus an increased mobility $B$), compared
to the pinned state $K<K_c$, see Fig.~\ref{fig:AubryFig}.

The driven FK model was usefully employed to describe the onset of sliding
of a crystalline contact \cite{Hammerberg98}, even though this model cannot
describe rigorously the experimentally-significant real-life plastic
deformations of the contact.

\begin{figure}
\centerline{
  \includegraphics[width=0.48\textwidth,clip=]{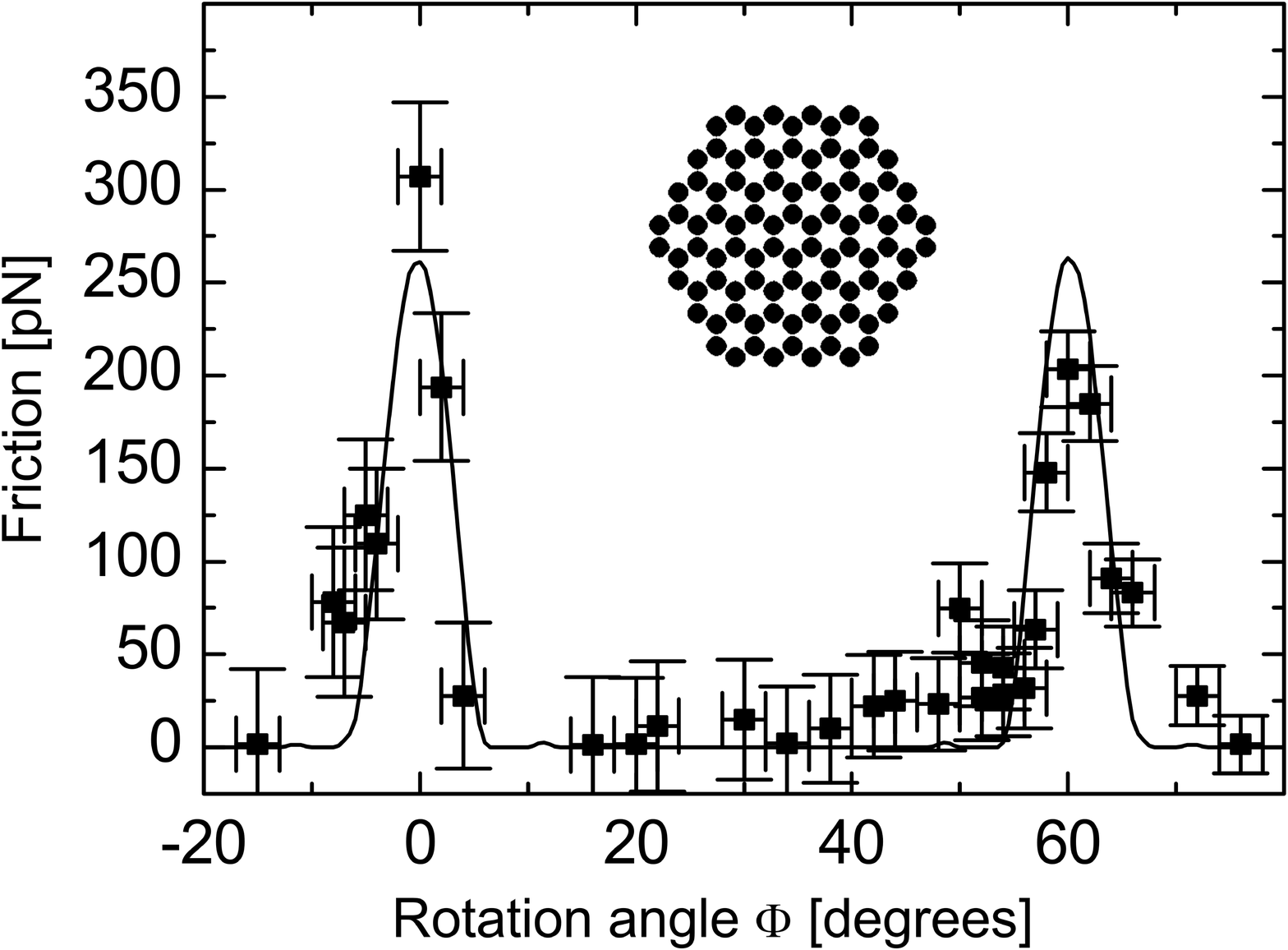} }
\caption{\label{fig:flakesuperlub}
  Data points: the average friction force as a function of the
  tip-substrate relative rotation angle measured in
  Ref.~\cite{Dienwiebel04}.
  Solid curve: the friction force computed from a generalized PT-like model
  based on the drawn 96-atom hexagonal flake.
  Adapted from Refs.~\cite{Dienwiebel04,Verhoeven04}, Copyright (2004) by
  The American Physical Society.
}
\end{figure}

Experimentally, superlubricity has been studied for a graphite flake
sticking to the tip of an atomic force microscope (AFM) sliding over an
atomically flat graphite surface \cite{Dienwiebel04, Verhoeven04,
  Dienwiebel05}.
Extremely weak friction forces of less than $50$~pN were detected in the
vast majority of the relative flake-substrate orientations, namely those
orientations generating incommensurate contacting surfaces, see
Fig.~\ref{fig:flakesuperlub}.
Stick-slip motion, associated with a much higher friction force (typically
$250$~pN), was instead found in the narrow ranges of orientation angles
where the flake-substrate contact was commensurate.

The above discussion ignores temperature, assuming so far $T=0$.
At nonzero temperature, the sliding-friction response of the FK model
\cite{Strunz98a,Strunz98b} requires of course the addition of a thermostat (see Sec.~\ref{linear:sec}).
The common choice of a Langevin thermostat for example simulates all
dissipation mechanisms through a viscous force $-m \gamma \dot x_i$, and
includes fluctuations by the addition of Gaussian random forces whose
variance is proportional to temperature $T$, as sketched in
Sec.~\ref{fluctdiss:sec}.
At $T>0$, thermal fluctuations can always overcome all sorts of pinning and
will thus initiate sliding by nucleation of mobile defects even in the
fully commensurate (pinned) condition, see Fig.~\ref{kinks}.

More generally, in the FK model the dimensionless coverage $\theta
=a_b/a_c$ plays a central role, because it defines the concentration of
``geometrical kinks" close to $\theta =1$ and of "superkinks" which arise
when $\theta$ deviates slightly from a background commensurate pattern that
is not $\theta =1 $, but rather a rational $\theta_0=p/q$, with $p$ and $q$
mutually prime integers.
If $\theta$ is only slightly different from $\theta_0$, the GS of the FK
model consists of extended domains with the commensurate pattern
associated to $\theta_0$, separated by superkinks (super-antikinks), in
the form locally mismatched regions of compression (expansion) relative
to $\theta_0$.

\begin{figure}
\centerline{
\includegraphics[width=0.48\textwidth,clip=]{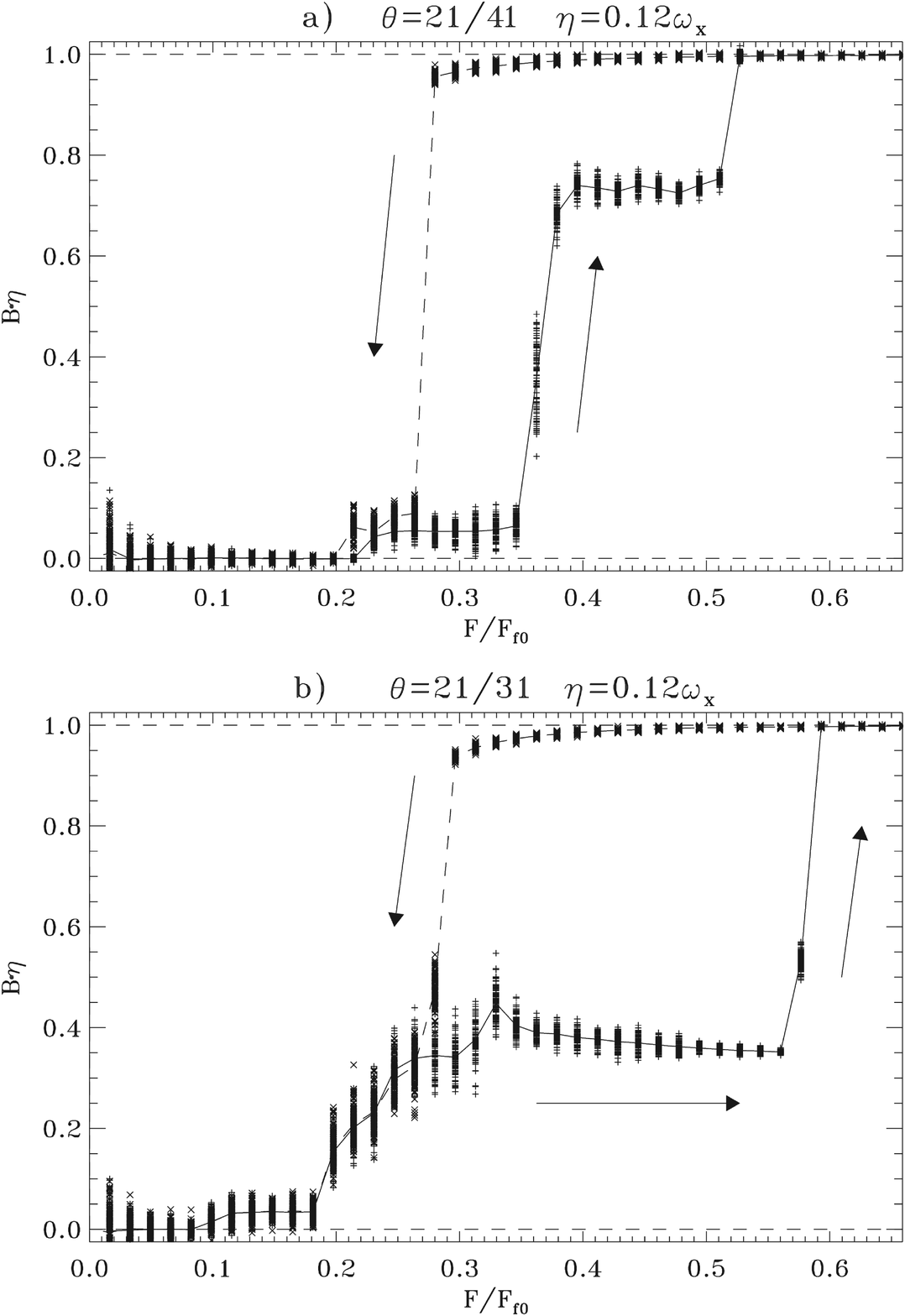}
}
\caption{\label{fig:Th-01}
  The mobility $B= v/F$ as a function of the driving force $F$ for the
  underdamped ($\gamma =0.12$) FK model with exponential interaction
  ($K_{\rm eff}=0.58$)
  (a)~for $\theta =21/41$ (superkinks on the background of a $\theta_0
  =1/2$ structure), and
  (b)~for $\theta =21/31$ (superkinks on the background of the complex
  $\theta_0 =2/3$ structure).
  From Ref.~\cite{Paliy97}, Copyright (1997) by The American Physical
  Society.
}
\end{figure}

The above concepts of pinning or superlubricity apply for an infinitesimal
applied force.
Additional interesting physics arises at finite force.
By increasing the driving force, a FK model with a pinned GS (either
commensurate, or incommensurate but past the Aubry transition) is known to
show a hierarchy of first-order dynamical phase transitions, starting from
the completely immobile state, passing through several intermediate stages
characterized by different running states of the kinks, to eventually reach
a totally running state.
Consider, for example, the ratio $\theta=21/41$: initially the mass
transport along the chain is supported by superkinks constructed on top of
the background $\theta_0=1/2$ structure.
Since the average superkink-superkink distance $41 \, a_b$ is large, they
interact weakly, and the atomic flow is restricted by the need for these
rarefied superkinks to negotiate their PN barriers (see
Fig.~\ref{fig:Th-01}a).
For larger driving $F$, the effective PN barriers are tilted and lowered
(in analogy to barriers of the corrugation potential), producing an
increased single-kink mobility $B$.
As a result, the zero-temperature transition from the locked state ($B=0$)
to the running state takes place at the force $F \approx F_{tk} = C \pi
\varepsilon_{\rm PN}/a_b$, where the factor $C \sim 1$ depends on the shape
of the PN potential.
In terms of the dimensionless superkink concentration $\theta_k=1/41$, the
mobility becomes $B\simeq \theta_k B_f$.

Beyond $F_{tk}$, further possibilities depend on the damping coefficient
$\gamma$.
At very small damping, $\gamma < 0.05$, the driven model transition leads
directly into the fully running state, because running superkinks
self-destroy soon after they start to move, causing an avalanche, thus
driving the whole chain to a total running state similar to that shown at
the right side of Fig.~\ref{kinks}.
When the dissipation rate is larger, $\gamma > 0.05$, one instead finds
intermediate stages with stable running superkinks, see
Fig.~\ref{fig:Th-01}.
The mechanism for a second rapid increase of the mobility after depinning
depends again on the value of $\gamma$ (for details see
Refs.~\cite{Braunbook, Paliy97, Braun98b}).
In between the initial superkink-sliding stage and the fully running state,
a sort of ``traffic-jam'' intermediate regime may emerge \cite{Braunbook}.

A qualitatively similar picture was confirmed also for different and more
complex kink patterns, such as that shown for the $\theta = 21/31$ example
in Fig.~\ref{fig:Th-01}b \cite{Paliy97}.
In this case, the GS consists of domains of the $\theta_0=2/3$ commensurate
structure, separated by superkinks at an average spacing $30\, a_b$.
Even the $\theta=2/3$ pattern  could  itself be viewed as a dense array
of trivial kinks constructed on top of the simple $\theta_0=1/2$ background
structure.
The force dependence of the mobility $B(F)$ bears a trace of this double
nature, with a state of running superkinks preceding a state of running
kinks.
For not too small $\gamma$ therefore the mobility $B$ increases in two
distinct steps as the driving force is increased.
A first step, at $F = F_{sk} \approx 0.08 \, F_0$ (here $F_0 = \pi U_0/a_b$
defines the depinning force for the fully commensurate model $\theta=1$)
occurs when the superkinks begin to slide; then a second step, at $F =
F^{\prime}_{tk} \approx 0.18 \, F_0$, occurs in correspondence to the
unpinning of the trivial kinks.

\subsection{Extensions of the FK Model}
\label{extensionsFK:sec}

Several extensions of the FK model have been proposed to describe a broad
range of frictionally relevant phenomena.
Most of these generalizations involve modifications of either the
interactions or the system dimensionality.
To address more realistic systems, anharmonicity of the chain interatomic
potential has been studied in detail.
The resulting features include mainly new types of dynamical solitons
(supersonic waves), a modification of the kink-kink interaction, the
breaking of the kink-antikink symmetry, and even the possibility of a chain
rupture associated to the excessive stretching of an antikink
\cite{Braunbook}.
The large kink-antikink asymmetry consistent with friction experiments in
layers of repulsive colloids \cite{Bohlein12} was attributed to the strong
anharmonicity in the colloid-colloid interaction
\cite{Vanossi12PNAS,Hasnain13}.
The essence of this asymmetry is the same as that between the physical
parameters of a vacancy and those of an interstitial.

Research has addressed also substrates with a complex corrugation pattern
\cite{Vanossi03, Remoissenet84}, including quasiperiodic \cite{Vanossi00,
  vanErp99} and random/disordered corrugation profiles \cite{Cule96,
  Cule98, Guerra07}.
Modifications from the plain FK model may generate qualitatively different
types of excitations, e.g.\ phonon branches and kinks of different kinds,
as well as modifications in the kink-antikink collisions.
At small driving force, where the dynamics and tribology are dominated by
moving kink-like structures, different sliding modes appear.

The Frenkel-Kontorova-Tomlinson (FKT) model \cite{Weiss96, Weiss97}
introduces an harmonic coupling of the sliding atomic chain to a driving
support, thus making it possible to investigate stick-slip features in a 1D
extended simplified contact.
The FKT framework provided the ideal platform to investigate the
tribological consequences of combined interface incommensurability,
finite-size effects, mechanical stiffness of the contacting materials, and
normal-load variations \cite{Igarashi08,Kim09}.

Important generalizations involving increased dimensionality compared to
the regular FK model bear significant implications for tribological
properties such as critical exponents, size-scaling of the friction force,
depinning mechanisms, and others.
In particular 2D extensions of the FK model \cite{PerssonBook, Braunbook}
have been applied to the modeling of the (unlubricated) contact of two
crystals.
Such is the case, for example, in quartz-crystal microbalance (QCM)
experiments, where single-layer adsorbate islands are made slide over a
crystalline substrate \cite{Krim88}.
Another case is that of recent experiments carried out with 2D monolayers
of colloids driven over a laser-generated optical lattice \cite{Bohlein12,
  Vanossi12, Vanossi12PNAS, Bohlein12PRL, Hasnain13, Hasnain14,
  MandelliPRL15, MandelliPRB15}.

Interestingly, the 2D Aubry transition of incommensurate colloids was
shown by Mandelli {\it et al.}~\cite{MandelliPRB15} to be of first
order, rather than of second order as in 1D.
As a consequence in 2D the free sliding and the pinned phases retain
local stability for a range of parameters that extends beyond the
transition point, a point where the total energy has a crossing
singularity instead of a smooth stiffness dependence as in 1D.
It is likely, although not proven to our knowledge, that the 2D FK should
possess a first-order Aubry transition too.

Among generalized 2D FK models we recall the two coupled FK chains
\cite{Roder98}, the ``balls and springs'' layer of particles linked in 2D
by harmonic springs and moving in a 2D periodic corrugation potential, the
scalar anisotropic 2D FK model consisting of a coupled array of 1D FK
chains, the 2D vector anisotropic model (namely the zigzag FK model where
the transverse motion of atoms is included
\cite{Braun98c,Braun95a,Braun95b,Braun94,Braun93,Braun91}), the 2D vector
isotropic FK model \cite{Persson93a, Persson93b, Persson93c, Persson94},
and finally the 2D tribology model \cite{Hammerberg98, Mikulla98} (see also
Ref.~\cite{Braunbook} and references therein).

These approaches which generalize the FK model have been of use for the
study of the transient dynamics at the onset of sliding.
Capturing these transient phenomena is often highly nontrivial in fully
realistic MD simulations (see e.g.\ Ref.~\cite{Braun01b}).
An interesting example of such a transient is the depinning of an atomic
monolayer driven across a 2D periodic substrate profile of hexagonal
symmetry \cite{Braun01b}.
The formation, by nucleation, of an island of moving atoms in a ``sea'' of
quasi-stationary particles mediates the transition from the locked to
the running state.
The moving island expands rapidly along the direction of the driving
force, and grows at a slower rate in the orthogonal direction.
Within the island, the 2D crystal retains its approximate ordered hexagonal
structure, thanks to its stiffness supported by the intra-atomic forces.
As a result, at the onset of depinning, the model exhibits regions of
almost perfect hexagonal-lattice order delimited by a closed boundary
of dislocations.

Kinks in 1D and dislocation lines in 2D exhibit peculiar tribological
properties.
In noncontact experiments, an oscillating AFM tip was seen to dissipate
significantly more when hovering above a dislocation line of a
incommensurate adsorbates than above in-registry regions \cite{Maier08}.
The larger softness and mobility of the dislocation regions
\cite{Gauthier00, Bennewitz00, Loppacher00, Hoffmann01} accounts for this
effect.
An explicit demonstration of this mechanism was carried out by the study of
a mismatched FK chain, whose dynamics was forced and simultaneously probed
by an oscillating localized model tip \cite{Negri10}.
This approach illustrates the ability of the FK model to capture the local
modifications of the dissipation properties.
In contrast, if retardation effects related to the finite speed of sound
across a material need to be taken into account, more sophisticated models
are called for.

\subsection{Quantized sliding velocity}

The investigation of systems {\em confined} between two shearing sliders,
such as single particles \cite{Rozman96a, Rozman96b, Muser02} or harmonic
chains \cite{Rozman97, Rozman98b, Braun05} embedded between competing
periodic potentials, have led to the discovery of several nonlinear
tribological phenomena involving either stick-slip dynamics or the
formation of peculiar ``synchronized'' sliding regimes
\cite{Drummond01,Drummond03}.

\begin{figure}
\centerline{
\includegraphics[width=0.48\textwidth,clip=]{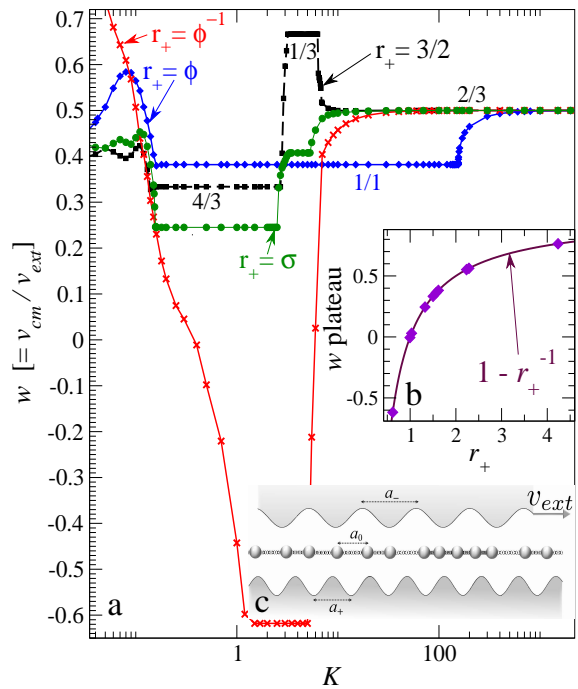}
}
\caption{\label{velcm:fig}
(a)~The ratio $w=v_{\rm cm}/v_{\rm ext}$ of the mean drift velocity of the
  chain to the top advancing speed, as a function of the chain stiffness
  $K$ for several lattice spacing ratios $(r_+,r_-)$, with
  $r_{\pm}=a_{\pm}/a_0$: commensurate $(3/2,9/4)$, golden mean (GM)
  $(\phi,\phi^2)$ ($\phi \simeq 1.6180\dots$), spiral mean (SM)
  $(\sigma,\sigma^2)$ ($\sigma \simeq 1.3247\dots$), and
  $(\phi^{-1},\phi)$.
  (b)~As a function of the length ratio $r_+$, the main plateau speed $w$
  of many calculations is seen to coincide with $1-r_+^{-1}$.
  (c)~A sketch of the model.
  Adapted from Ref.~\cite{Vanossi06}, Copyright (2006) by The American
  Physical Society.
}
\end{figure}

The FK model can be generalized with the addition of a second, different
sinusoidal corrugation potential, as sketched in Fig.~\ref{velcm:fig}c.
When the second potential is spatially advanced relatively to the first as
a function of time, the model realizes the simplest idealization of a
slider-solid lubricant-slider confined geometry.
In this extended model the lattice mismatch was shown to generate peculiar
and robust ``quantized'' sliding regimes \cite{Braun05,Santoro06,
  Vanossi06, Cesaratto07, Manini07extended, Vanossi07Hyst, Vanossi07PRL,
  Manini07PRE, Manini08Erice}, where the chain deformations are
synchronized to the relative motion of the two corrugations, in such a way
that the chain's (i.e.\ the solid lubricant's) average drift velocity
acquires nontrivial fixed ratios to the externally-imposed sliding
velocity.
Specifically, the ratio of the lubricant speed to that of the slider
$w=v_{\rm cm}/v_{\rm ext}$ remains locked to specific ``plateau'' values
across broad ranges of most model parameters, including the potential
magnitude of the two sliders, the chain stiffness (see
Fig.~\ref{velcm:fig}a), the dissipation rate $\gamma$, and even the
external velocity $v_{\rm ext}$ itself.
The speed ratio is ultimately determined by geometry alone: $w=1-r_+^{-1}$,
where $r_+ = a_+/a_0$ is the incommensurability ratio between the chain
spacing $a_0$ and that, $a_+$, of the closest slider \cite{Vanossi06}.
The plateau mechanism is the fact that the kinks formed by the mismatch of
the chain with one slider (the slider whose spatial periodicity is closest
to that of the chain), are rigidly dragged at velocity $v_{\rm ext}$ by the
other slider.
The kink density being geometrically determined and lower than the chain
density implies that the overall velocity ratio shares exactly the same
properties.
The exactness of the velocity plateaus implies a sort of ``dynamical
incompressibility'', identically zero compliance to perturbations trying to
modify $v_{\rm cm}$ from its quantized-plateau value.
This robustness of the plateaus can be demonstrated e.g.\ by adding a
constant force $F_{\rm ext}$, pushing all particles in the chain:
as long as $F_{\rm ext}$ is small enough, it does perturb the dynamics of
the velocity-plateau attractor, but not the value of $v_{\rm cm}$.
Eventually, above a critical force $F_c$, the driven model abandons the
plateau dynamics.
This transition, explored by increasing the external driving force $F_{\rm
  ext}$, exhibits a broad hysteresis, and shares many features of the {\em
  static}-friction depinning transition, except that here it takes place
``on the fly'' \cite{Vanossi07PRL, Manini07extended}.
Disregarding details, this transition is then formally equivalent to the
standard Aubry depinning transition \cite{AubryLeDaeron, Peyrard83}, with
the moving kinks of the lubricant-substrate interface taking here the role
of particles.
The robustness of the quantized plateau stands even after replacement of
the sinusoidal corrugation potential of Eq.~\eqref{FKHamil} with a deformed
profile: the Remoissenet-Peyrard non-sinusoidal potential even extends the
velocity plateau in the space of model parameters \cite{Woulache13}.

\begin{figure}
\centerline{
\includegraphics[width=0.6\textwidth,clip=]{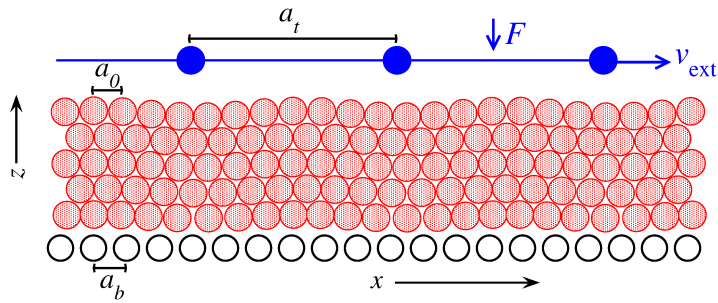}
}
\caption{\label{model_quantized:fig}
  The model consists of the rigid top (solid circles) and bottom (open)
  layers, plus one or several lubricant layers confined in between.
  The lattice spacing are $a_{\rm t}$ and $a_{\rm b}$ and (on average)
  $a_{\rm 0}$, respectively.
  The top layer advances at an externally imposed $x$-velocity $v_{\rm ext}$.
  From Ref.~\cite{Castelli08Lyon}, Copyright (2008) IOP Publishing Ltd.
}
\end{figure}

\begin{figure}
\centerline{
\includegraphics[width=0.56\textwidth,clip=]{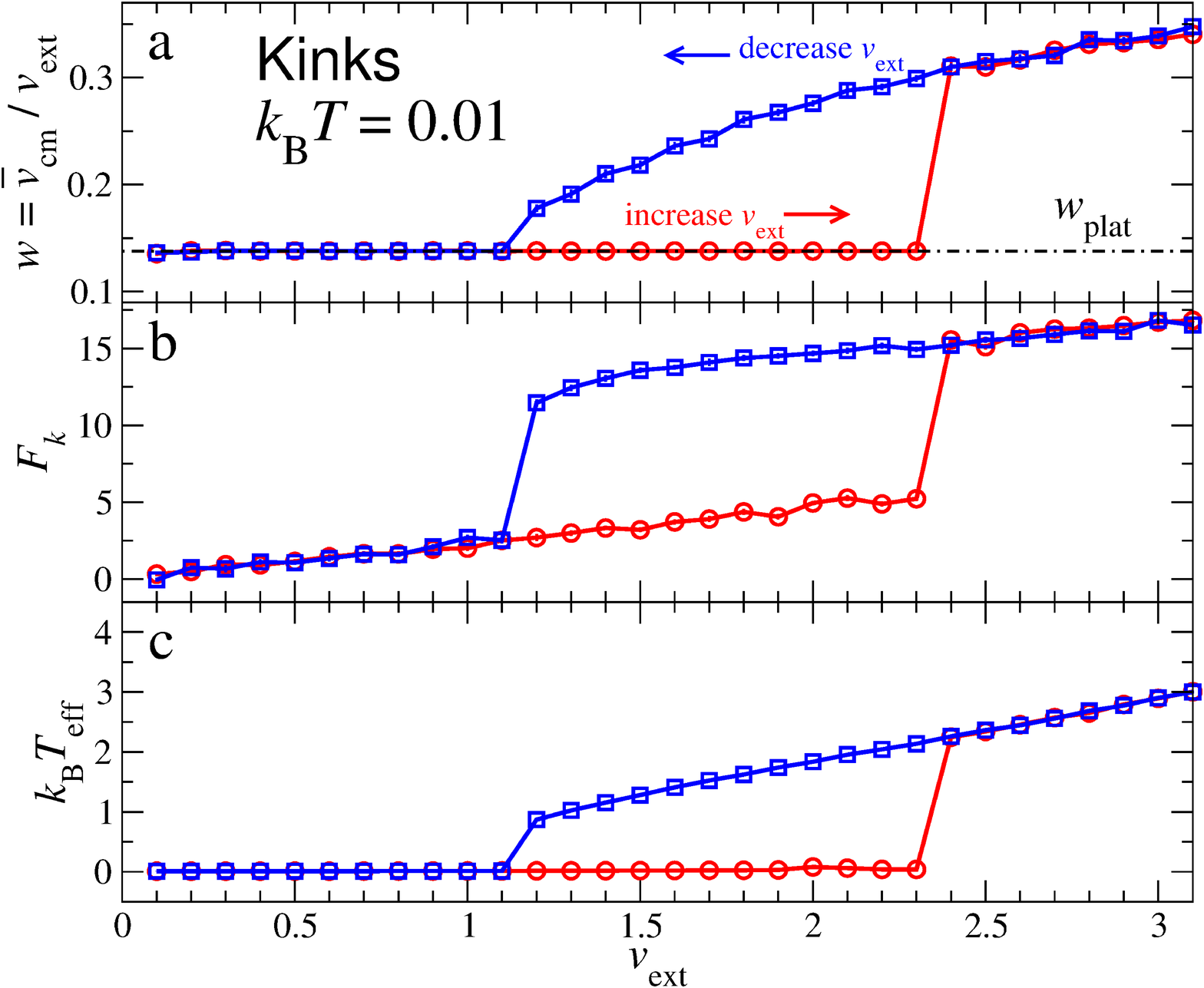}
}
\caption{\label{hyst-diss-kink:fig}
  For the model of Fig.~\ref{model_quantized:fig}, as functions of the
  adiabatically increased (circles) or decreased (squares) driving velocity
  $v_{\rm ext}$, the three panels report:
  (a)~the average velocity ratio $w=v_{\rm cm}/v_{\rm ext}$;
  (b)~the average friction force $F_k$ needed to make the top substrate
  advance;
  (c)~the lubricant temperature, evaluated based on the average kinetic
  energy in the reference frame of the instantaneous lubricant center of
  mass.
  The lubricant and the sliders are Lennard-Jones atoms, and temperatures,
  including the thermostat temperature $k_{\rm B} T =0.01$, are expressed
  in units of the depth of the interaction potential well.
  Adapted from Ref.~\cite{Castelli09}, re-used under permission of a
  Creative Commons license.
}
\end{figure}

The quantized sliding regime of the crystalline solid lubricant was also
investigated in a significantly less idealized 2D model, including the
perpendicular degree of freedom \cite{Castelli08Lyon}.
MD simulations carried out for a monolayer or multilayer lubricant film
where atoms interacting via Lennard-Jones potentials can also move
perpendicularly to the sliding direction, as sketched in
Fig.~\ref{model_quantized:fig} showed quantized plateaus in this case too.
These plateaus were shown to be resilient against a variations in the
loading forces across a broad range, against thermal fluctuations, and also
against the presence of quenched disorder in the substrates.
This quantized sliding state was also characterized by significantly lower
values of kinetic friction $F_k$ (the average force needed to maintain the
advancement of the top slider) \cite{Castelli09}, than the regular
non-quantized regime, see Fig.~\ref{hyst-diss-kink:fig}.

Quantized sliding has been again demonstrated more recently in a 3D model
where the lubricant is represented by a layer of Lennard-Jones atoms
\cite{Vigentini14}.
The quantized-sliding state and its boundaries were fully characterized in
the special case of perfectly-aligned crystalline layers.
We note however there are reasons to expect that incommensurately
mismatched epitaxial layers could relax to a mutually rotated alignment
\cite{Novaco77,ManiniBraun11}.
The quantized-sliding state in such rotated arrangements is the subject of
active investigation.
More generally, no experimental observations of the quantized sliding
predicted for solid lubricants has appeared so far.
Layered systems such as graphene and BN appear to offer a good opportunity
for the future study of these curious phenomena.

%=========================================================================
\section{Molecular Dynamics Simulations}
\label{MD:sec}

The simple models considered above yielded precious qualitative and often
semi-quantitative understanding of several features of friction.
To address subtler physical behavior in specific systems, it is
nevertheless necessary and desirable to include atomistic structural and
mechanical details of the interface.
MD simulations can help make an inroad such detail, also offering a level
of detail that can in some instances replace experiment.

\begin{figure}
\centerline{
\includegraphics[height=0.3\textwidth,clip=]{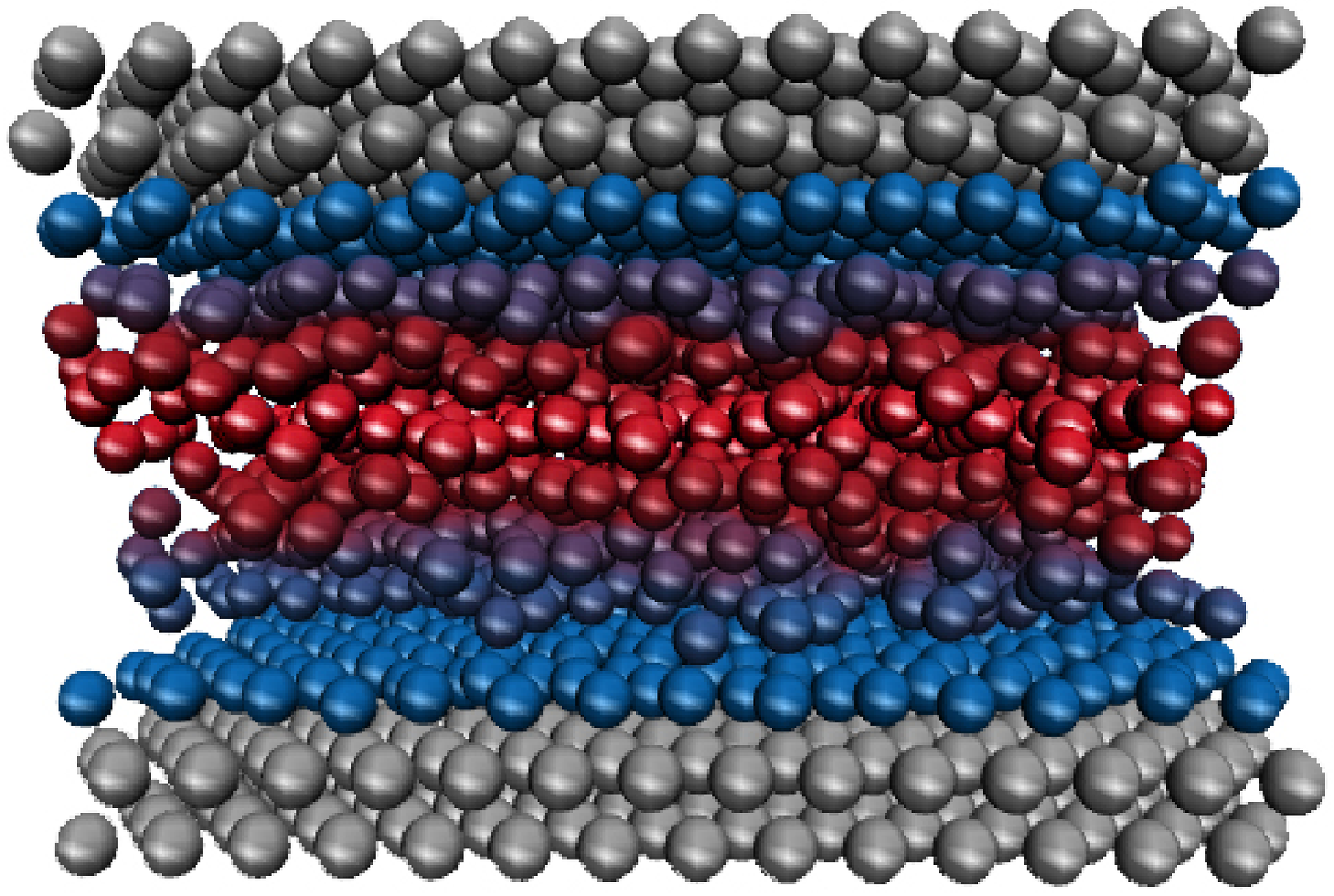}
\hfill
\includegraphics[height=0.3\textwidth,clip=]{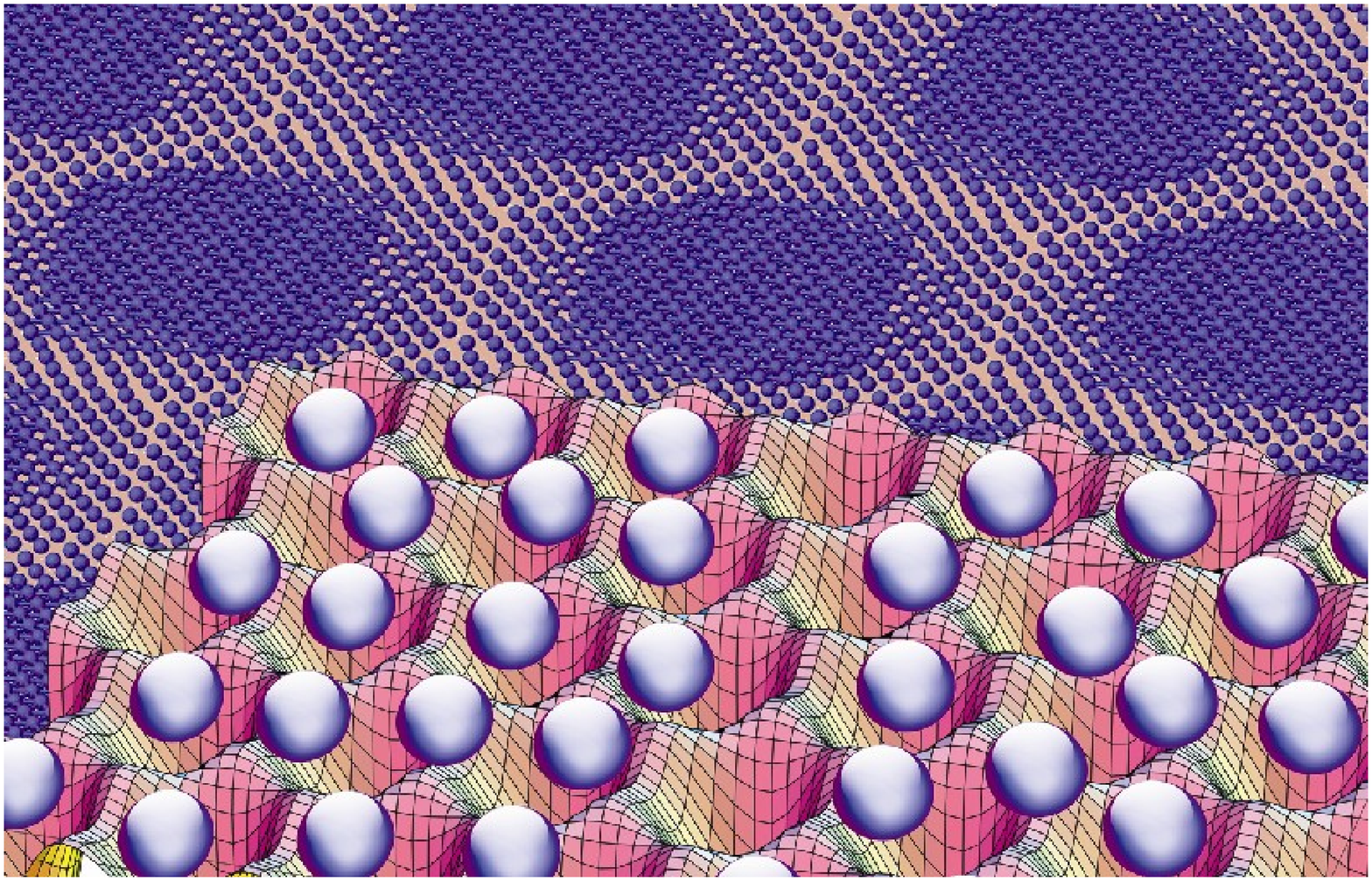}
}
\caption{\label{simulation:fig}
Two sketches of typical MD simulations.
Left: the melting induced by shearing at a boundary-lubricated interface
\cite{Vanossi13}.
Right: simulations of a monolayer of colloids interacting with each other
and with a periodic corrugated potential produced by a laser interference
pattern as in Ref.~\cite{Vanossi12PNAS}.
}
\end{figure}

Thanks to advances in computing algorithms and hardware, recent years have
witnessed a remarkable increase in our ability to simulate tribologic
processes in realistic nano-frictional systems, and obtain detailed
microscopic information.
A MD simulation is {\it de facto} a controlled computational experiment,
where the overall atomic dynamics is provided by the numerical solutions of
suitably generalized Newton's equation of motion relying on interatomic
forces derived by specific realistic interparticle-interaction potentials.
Tribological simulations require a careful selection of the geometric
arrangement of the sliding interface, e.g.\ as in
Fig.~\ref{simulation:fig}, and of the applied boundary conditions.

Influential review articles \cite{Robbins01, Muser06} cover the atomistic
MD simulation of friction, with a focus on technical aspects such as the
construction of a realistic interface, suitable techniques for the
application of load, shear, and temperature control.
The simplest approach to temperature control, namely adding a Langevin
thermostat to Newton's equations \cite{Frenkel-Smit96}, has been adopted
broadly, but more refined approaches have been proposed and adopted for
friction simulations, as discussed in Sect.~\ref{thermostat:sect} below.
Physically relevant quantities, including the average friction force, the
slider and the lubricant mean velocities, several correlation functions,
and the heat flow can be evaluated numerically by carrying out suitable
averages over the model dynamics of a sliding interface, as long as it is
followed for a sufficiently long time.
The modeling of friction must first of all address correctly ordinary
equilibrium and near-equilibrium phenomena, where the
fluctuation-dissipation theorem (Sec.~\ref{linear:sec}) governs the smooth conversion of mechanical
energy into heat, but most importantly it must also deal with inherently
nonlinear dissipative phenomena such as instabilities, stick-slip, and all
kinds of hysteretic response to external driving forces, characteristic of
non-equilibrium dynamics.

The choice of realistic interatomic forces is often a major problem.
Indicating with $U\{R_1, R_2, ...R_N\}$ the total interaction energy as a
function of all atomic coordinates $\{R_i\}$, the force on atom $i$ is $F_i
= - \nabla_{R_i} U$, fully determined in terms of $U$.
Unfortunately, the adiabatic energy $U$ results from the solution of the
quantum ground state of the electrons --- a practically complicated problem
whose quantitative outcome may moreover be of uncertain quality.
The reason why {\it ab-initio} MD, e.g.\ of the Car-Parrinello type
\cite{Car85}, is not generally used in sliding friction is that it can
neither handle large systems, exceeding few hundreds of atoms, nor run for
tribologically significant duration, usually in excess of $\sim 1$~ns.
On the other hand, the physical situations where a first-principles
description of interatomic forces is mandatory are not too common in the
frictional phenomena studied so far.
As a consequence, most MD models for friction rely on more or less refined
interatomic ``force fields'', ranging from sophisticated energy surfaces
modeled on calculations at {\it ab-initio} density-functional or
tight-binding level \cite{Xu92,Bonelli09}, to empirical distance- and
angle-dependent many-body classical potentials, to basic pairwise
potentials (e.g.\ Morse or Lennard-Jones), to the simplest elastic-springs
models, which represent generalizations of the FK model.
Concretely, the scientific literature documents many realistic force
fields, ready to address several classes of materials and their combinations
\cite{Finnis03,Rafii-Tabar04}.
While these force fields allow qualitative atomistic simulations of
tribological systems, their limitations often prevent quantitative
accuracy.
In particular, in the course of such a violent frictional process as wear,
atoms are likely to modify their chemical coordination and even their
charge state: phenomena and radical chemical changes usually impossible to
describe with empirical force fields.
Mechanochemistry and tribochemistry are time-honored areas \cite{Gilman96,
  Fischer88} offering obvious examples where empirical force-fields would
fail, and simulations must by necessity be conducted by
electronic-structure based first-principles methods.
Also, even if for a specific system, element or compound, a satisfactory
force field has been arrived at, the mere replacement of one atomic species
with another one generally requires a complete and usually difficult
re-parameterization of the whole force field.
As a result, quantitatively accurate nanofrictional investigations remain a
substantial challenge because of opposite limitations in the use of first
principles versus empirical force fields.
A promising compromise could possibly be provided by the so-called reactive
potentials \cite{Stuart00, Brenner01, Adri01}, capable of describing some
chemical reactions, including interface wear with satisfactory
computational efficiency in large-scale atomic simulations, compared to
semi-empirical and first-principles approaches.

Retardation effects due to the finiteness of the speed of sound are usually
irrelevant in slow-speed experiments ($v<1$~mm/s).
For larger speed, retardation effects related to the finite speed of sound
can be taken into account explicitly in MD modeling, provided that rigid
layers are either omitted or introduced with special care.
Other effects of nonlocality in time, such as retardation due to the finite
speed of light \cite{VolokitinRMP07,Volokitin11} are usually omitted in the MD
force fields altogether, as they lead to negligible corrections in all
conditions where sliding involves a proper material contact.
Such retardation effects do play a role in noncontact geometries, such as
in experiments probing lateral Casimir forces \cite{Chiu10}, whose strength
can become relevant at large sliding speed \cite{Zhao12}.

%==================================================================
\subsection{Thermostats: the Dissipation of Joule Heat}
\label{thermostat:sect}

As we already mentioned above in Sec.~\ref{linear:sec},
any kind of sliding friction involves mechanical work, some of which is
then transformed into heat (the rest going into structural transformations,
wear, etc.).
The heat is then transported away by phonons (and electrons in the case of
metallic sliders) and eventually dissipated to the environment.
Likewise, all excitations generated at the sliding interface in simulations
should be allowed to propagate away from it, and to disperse in the bulk of
both sliders.
Instead, due to the small simulation size, this energy may unphysically
pile up in the rather small portion of solid representative of the ``bulk''
of the substrates, where these excitations are scattered and back-reflected
by the simulation-cell boundary, instead of being properly dissipated away.
In order to prevent continuous heating and attain a steady state of the
tribological system, the Joule heat must then be removed at a steady rate.
In the FK and PT models, a viscous damping term $-m \gamma \dot x_i$,
Eq.~\eqref{eq_Langevin}, is generally introduced for this purpose.
In these minimal models however, the value of $\gamma$ is well known to
affect the dynamical and frictional properties, but there is unfortunately
no clear prescription for the choice of $\gamma$.
In MD atomistic simulations, the heat removal is often achieved by means
of equilibrium thermostats, e.g.\ Nos\'e-Hoover or Langevin, see
Sec.~\ref{fluctdiss:sec}.
In this way however an unphysical energy sink is spread throughout the
simulation cell.
As a result, the atoms at the interface fail to follow their actual
conservative trajectories, but evolve through an unphysically damped
dynamics, with unknown and generally undesired effects on the overall
tribological properties \cite{Tomassone97}.
In order to address and mitigate this problem, modifications of the
equations of motion for the atoms inside the microscopically small
simulation cell were proposed with the target of reproducing the frictional
dynamics of a realistic macroscopic system, after the integration of extra
``environment'' variables.
One possible approach is the application of Langevin equations with a damping
coefficient that changes as a function of the position and velocity of each
atom in the lubricant, in accordance with the dissipation known for the
atoms adsorbed on a surface \cite{Braun01a}.
This method involves modifying the standard Langevin equations
\cite{Braun02b}.
Another approach to improve the simulation of dissipation within blocks in
reciprocal motion requires modifying the damping term \eqref{eq_Langevin}
to a form
\begin{equation}\label{eq_Langevin_cm}
  \vec{f}_{{\rm damp}\,j} =
  - m \gamma (\dot{\vec{r}}_j-\vec{v}_{\rm loc}) \,,
\end{equation}
where $\vec{v}_{\rm loc}$ is the average center-mass velocity of the atoms
forming the sliding block to which particle $j$ belongs locally
\cite{Reguzzoni10,Pierno15}.
Another more rigorous, physically appealing approach is the
recently-implemented dissipation scheme, drawing on earlier, long-known
formulations \cite{Magalinski59,Rubin60,Zwanzig73} and subsequent
derivations \cite{Li07,Kantorovich08a,Kantorovich08b}, describing the
correct embedding of the Newtonian simulation cluster inside a larger heat
bath made of the same material.
Upon integrating out the heat bath degrees of freedom, atoms in the
boundary layer that borders between the cluster and the heat bath are
subjected to additional non-conservative and non-Markovian forces that
mimic the surrounding bath through a so-called memory kernel.
An approximate but very practical scheme replaces this memory kernel by a
simple viscous damping $\gamma$, here applied exclusively to atoms in the
boundary layer.
The magnitude of the parameter $\gamma$ is optimized variationally by
minimizing, with surprising accuracy, the energy reflected across the
boundary \cite{Benassi10,Benassi12}.
This dissipation scheme has been implemented recently in nanofriction
simulations where it was shown to improve greatly over other conceptually
and practically inadequate thermostats.

Besides the limitations of system size and simulation times that are
obvious and will be discussed later, there is another limitation concerning
temperature, that is rarely mentioned.
All classical frictional simulations, atomistic or otherwise, are only
valid at sufficiently high temperature.
They become in principle invalid at low temperatures where the mechanical
degrees of freedom of solids progressively undergo "quantum freezing", and
both mechanics and thermodynamics deviate from classical.
Unfortunately there is at the moment no available route to include
appropriately these quantum effects in dynamical and frictional
simulations.

%==================================================================
\subsection{Size and timescale issues}
\label{sizeissues:sec}

Each core of a present-day CPU executes $\sim 10^9$ floating-point operations
per second (FLOPS).
MD simulations usually benefit effectively of medium-scale parallelization.
Approximately linear scaling can be achieved up to $\sim 100$ cores, thus a
MD simulation can execute $\sim 10^{11}$ FLOPS routinely.
The evaluation of the forces is usually the most CPU-intensive part of a MD
simulation.
For each atom, depending on the force-field complexity and range, this
evaluation can require $\sim 10 - 10^2$ operations, or even more.
As a result, the number of time-integration steps $N_{\rm step}$ multiplied
by the number of simulated particles $N$, is $\sim N N_{\rm step}\simeq
10^{10}$ per computer runtime second.
Given that simulations of atomic-scale friction require time-steps in the
femtosecond region, a medium-size simulation involving $N=10^5$ simulated
particles, can advance at an estimated speed of $\sim 10^5$~fs each
real-life second, namely $\sim10^9~{\rm fs} = 1~\mu$s each simulation day.
Clearly, speed scales down for more refined force fields, and for larger
systems size, although this increase may be mitigated by a larger-scale
parallelization.

We can compare these estimates with typical sizes, duration, and speeds in
frictional experiments.
In macroscopic tribology experiments, sliding speeds often range in the
$0.1 - 10$~m/s region: each microsecond the slider would progress by $0.1$
to $10~\mu$m, namely $\sim 10^3 - 10^4$ lattice spacings of standard
crystalline surfaces.
In such conditions $10^{-3} - 10^{-2} ~\mu$s may suffice to generate a good
statistics of atomic-scale events, although it may still be insufficient to
address e.g.\ the diffusion of wear particles or additives in the
interface, or phenomena associated to surface steps and/or point defects.
By contrast, in nanoscale AFM experiments the tip usually advances at much
lower speeds $\simeq 1~\mu{\rm m/s}$: over a typical run it is possible to
simulate a tiny $\sim 1$~pm displacement, far too small to explore even a
single atomic-scale event, let alone averaging over a steady state.
For this reason, in all conditions where long equilibration times and/or
slow diffusive phenomena and/or long-distance correlations can be expected,
models should be preferred to realistic but expensive MD.
However, MD simulations can provide so much physical insight that they make
sense even if carried out at much higher speeds than in real-life AFM or
surface force apparatus (SFA) experiments: in practice, currently the
sliding speeds of most atomistic tribology simulations are in the $\sim
1$~m/s region.

Here however we should distinguish between static and kinetic friction,
and for the latter between smooth-sliding and stick-slip regimes.
Smooth kinetic friction generally increases with speed (velocity
strengthening), but sometimes decreases with increasing speed in certain
intervals.
In the former case, simulating smooth high-speed frictional sliding is not
fundamentally different from the real sliding at low speed, with
appropriate changes in frictional forces with $v$.
Velocity weakening conditions, alternatively, tend to lead to an intrinsic
instability of smooth sliding, which is therefore not often pertinent to
real situations.
As a result, for nanoscale systems, MD simulations is of value in the
description of smooth dry kinetic friction despite the huge velocity gap.
On the other hand, static friction -- the smallest force needed to set a
slider in motion -- is also dependent on the simulation time (a longer wait
may lead to depinning when a short wait might not), and generally dependent
on system size, often increasing with sub-linear scaling with the slider's
contact area.
To address this kind of behavior in MD simulations, it is often necessary
to resort to scaling arguments in order to extrapolate the large-area
static friction from small-size MD simulations \cite{Braun13, Pierno15}.

Returning to the simulation time problem, let us come to stick-slip in MD
simulation, and to the desirability to describe the stick-slip to
smooth-sliding transition as a function of parameters such as speed.
In AFM and SFA experiments, stick-slip and its associated
characteristically high friction and mechanical hysteresis tend to
transition into smooth sliding when the speed exceeds $\sim 1~\mu$m/s; in
contrast, in MD modeling the same transition is observed in the $\sim
1$~m/s region.
This 6-order-of-magnitude discrepancy in speed between experiments and
simulations is well known and has been largely discussed \cite{Braun02a,
  Luan04, BPBFV2005, Braun06} in connection with the effective mass
distributions and spring-force constants, that are vastly different, and
highly simplified in simulations.
Attempts to fill the time and speed gaps can rely on methods, such as
hyperdynamics, parallel-replica dynamics, on-the-fly kinetic Monte Carlo,
and temperature-accelerated dynamics which have been developed in the last
decades \cite{Voter02, Mishin07, Kim11}.
However, caution should generally be exerted in that some of these schemes
and methods are meant to accelerate the establishment of equilibrium but
not always to treat properly the actual frictional-loss mechanisms.
Concerning stick-slip friction, another problem is that, unlike
simulations, real experiments contain mesoscale or macroscale component
intrinsically involved in the mechanical instabilities of which stick-slip
consists.
Here the comforting observation is that stick-slip is nearly independent of
speed, so that so long as a simulation is long enough to realize a
sufficient number of slip events, the results may already be good enough
\cite{molshapeLin}.
One can even describe stick-slip friction adiabatically, e.g., from a
sequence of totally static calculations, where a periodic back-and-forth
sliding path is trodden, the area of hysteresis cycle generated by two
different to and from instabilities representing the friction
\cite{Oyabu06}.

A serious aspect of stick-slip friction which MD simulation is unable to
attack is ageing.
The slip is a fast event, well described by MD, but sticking is a long
waiting time, during which the frictional contact settles very slowly.
The longer the sticking time, the larger the static friction force
necessary to cause the slip.
Typically experiments show a logarithmic increase of static friction with
time \cite{Dieterich94}.
Rate and state friction approaches, widely used in geophysics
\cite{Ruina83}, describe phenomenologically frictional ageing, but a
quantitative microscopic description is still lacking.
Mechanisms invoked to account for contact ageing include chemical
strengthening at the interface in nanoscale systems \cite{Li11b}, and
plastic creep phenomena in macroscopic systems \cite{Heslot94}.
Contact ageing is observed also in other disordered systems out
of equilibrium, including glasses and granular matter.
In seismology finally, as will be discussed later, it is generally accepted
that ageing is responsible for aftershocks, as also shown by some
models \cite{BT2014}.

%==================================================================
\subsection{Multiscale Models}

If MD simulation may be satisfactory in nanoscale friction, it is clearly
not capable of describing mesoscale and macroscale tribology.
The insurmountable difficulties of the fully atomistic treatment of all
typical and large length scales that are responsible for the dynamical
processes in large scale systems has prompted in recent years increasing
efforts towards multiscale approaches.
The main idea is that, at a sufficient distance from the sliding interface,
continuum mechanics should describe all processes to a fair level of
approximation.
Finite-elements simulations of the continuum mechanics may provide a
practical model for the elastic and plastic deformations.
Using finite element methods, one can increase the coarse-graining level
while moving away from the sliding interface, thus keeping the
computational effort under control.
Several groups \cite{McGee07,Luan06} combined the MD description of the
sliding interface, where local deformations at the atomic length scale and
highly nonlinear phenomena occur, with a continuum-mechanics description in
the ``bulk'' regions where strains are continuous and small.
The main difficulty faced by this class of approaches is the correct choice
of the matching between the atomistic region with the continuum part
\cite{E09}.
Because at the continuum level the detail of lattice vibrations cannot be
represented in full, the matching conditions should at least minimize the
reflection of the acoustic phonons at the atomistic-continuum interface.
In other words, the matching should allow the transmission of sound
deformations in both directions with sufficient accuracy: this is necessary
for a proper disposal of the Joule heat into the bulk.

%==================================================================
\subsection{Selected Results of MD Simulations}

Simulations can provide direct insight in the dynamical processes at the
atomistic level, that are at the origin of friction, allowing a connection
of these microscopic facts with their macroscopic counterparts.
Case studies, in which the system is well describable by both experimental
and theoretical sides, are of extreme importance firstly to permit a
crosscheck between the two, and then to make use of simulations in order to
highlight particular aspects that cannot be accessed by experiments.
Here we summarize the main results of a few selected simulations sampled
from the expanding literature of friction simulation, certainly not
claiming an exhaustive review of the field.

\begin{figure}
\centering
\includegraphics[width=0.5\textwidth,clip=]{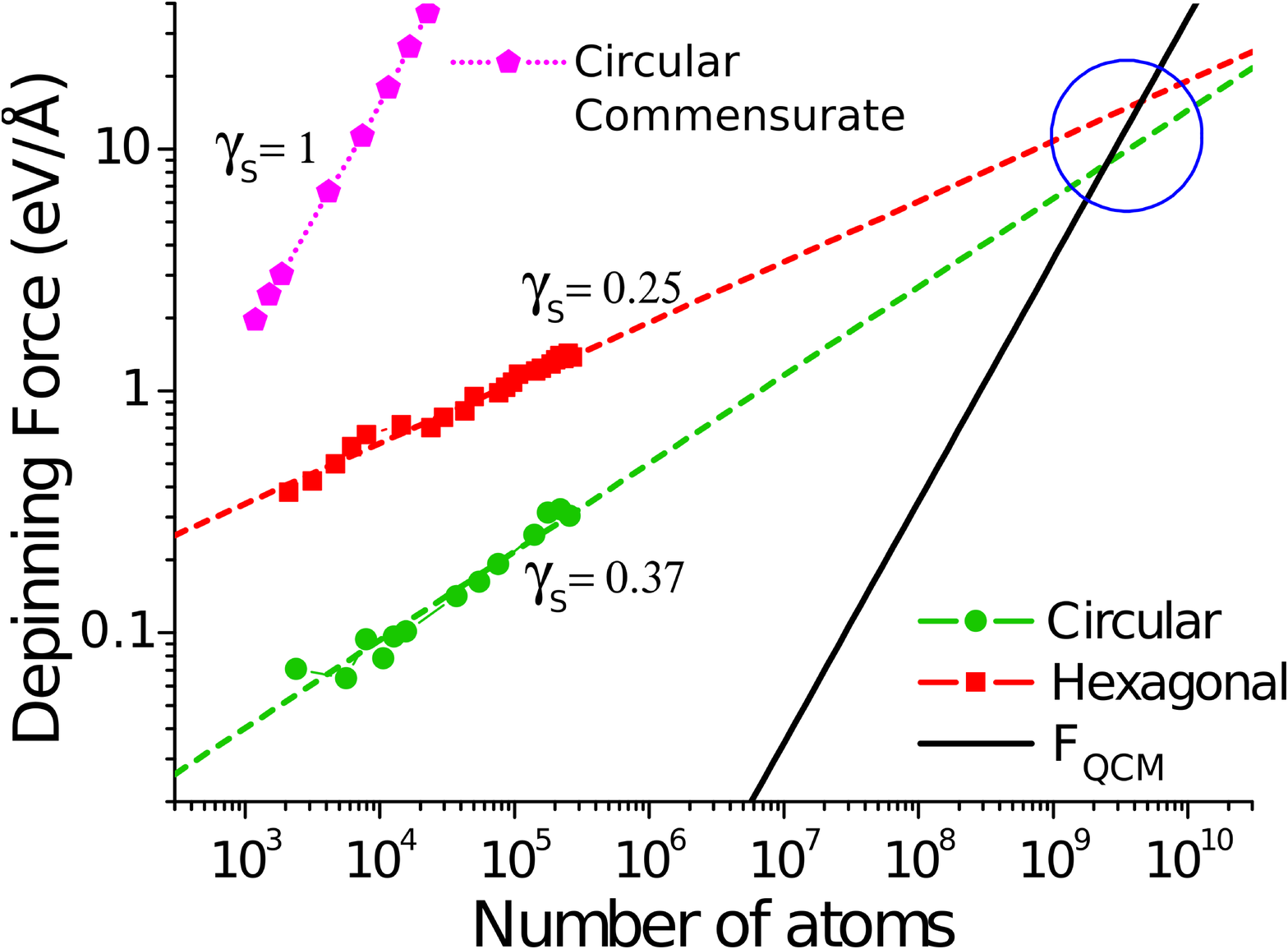}
\caption{\label{varini.fig2}
  Static friction scaling with contact area for simulated incommensurate Kr
  islands, of either circular or hexagonal shape, deposited on Pb(111).
  The static friction for the $\sqrt{3}\times \sqrt{3}$ commensurate
  arrangement and the QCM inertial force on the island are reported for
  comparison.
  Adapted from Ref.~\cite{Varini15}, Copyright (2015) by Royal Society of
  Chemistry.
}
\end{figure}

%ERIO:  QUI E' IL CASO DI SEPARARE I CASE STUDIES...PROVO...

\subsubsection{Rare-gas islands and metal clusters.}
The sliding of rare-gas overlayers deposited on metallic substrates at low
temperature has contributed much to the understanding of how friction
scales with the contact-area size, the substrate corrugation, and the
sliding velocity.
Rare-gas atoms condense into 2D solid islands showing a faceted-circular
shape, arranged on multiple layers at low temperatures or on a single layer
at diffusion-enabled temperatures \cite{Park99}.
The friction characteristics of these solid islands on the substrate,
resulting from inertial sliding, has been probed experimentally by QCM
apparatuses, revealing a complex interplay among friction, coverage, and
temperature \cite{Daly96,Pierno15}.
The rare-gas lattice spacing inside the island, generally incommensurate
but sometimes commensurate with that of the substrate, plays a very
important role in determining the pinning or free sliding that controls the
frictional behavior.
A generally overlooked aspect which has been highlighted only recently
\cite{Pierno15} is the larger thermal expansion coefficient of rare-gas
layers than that of a metal substrate, causing a temperature-dependent
lattice mismatch at the interface with possible incommensurate-commensurate
transitions.
Due to this mechanism, MD simulations have predicted the possible
appearance of static-friction peaks in the correspondence of a long-range
commensurate phase occurring at a particular temperature \cite{Varini15}.
In the case of periodic monolayers, a change in the lattice mismatch can be
also induced by an adhesion-driven densification of the adsorbate
\cite{Cieplak94}, again eventually encountering a commensurate phase
\cite{Daly96,Pierno15}.
Simulations of rare-gas incommensurate adsorbates, whose linear
substrate-induced misfit dislocations (``solitons'') must flow during
sliding, have revealed the role of their entrance in the depinning of the
island \cite{Varini15}, and of their dissipation through anharmonic
coupling to phonons, in kinetic friction \cite{Cieplak94}.
Finite-size effect are in this case of absolute relevance, effectively
generating (or enhancing) static friction through a pinning barrier arising
at the interface edge, which solitons must overcome to establish motion
\cite{Varini15}.
The edge-related origin of the pinning mechanisms implies that static
friction $F_s$ can grow with the island size $A$ at most as $F_s\propto
A^{1/2}$, i.e.\ as its perimeter $P \propto A^{1/2}$,
% NICK: Qui niente pigrechi, mica abbiamo detto la forma dell'isola...
if the pinning points were uniformly distributed along the island or
cluster edge.
As shown in Fig.~\ref{varini.fig2}, a different shape of the deposited
nano-object can generally lead to a different scaling
exponent.
Similar sublinear scaling exponents were identified in dynamic friction
experiments in which gold nanoclusters of variable size/shape were dragged
at low speed over a graphite substrate by an AFM tip \cite{Dietzel13}.
Scaling exponents of both the rare-gas island/metal surfaces
(theoretical) and dragged gold clusters/graphite (experimental) are in
the order of $1/4$.
This indicates that not all points at the boundary provide pinning with
equal efficiency.
A scaling close to $\sim A^{1/4}$ might rather indicate a random
efficiency of boundary points, whereby only $\sim P^{1/2}$ provide
effective pinning.

\begin{figure}
\centering
\includegraphics[width=\textwidth,clip=]{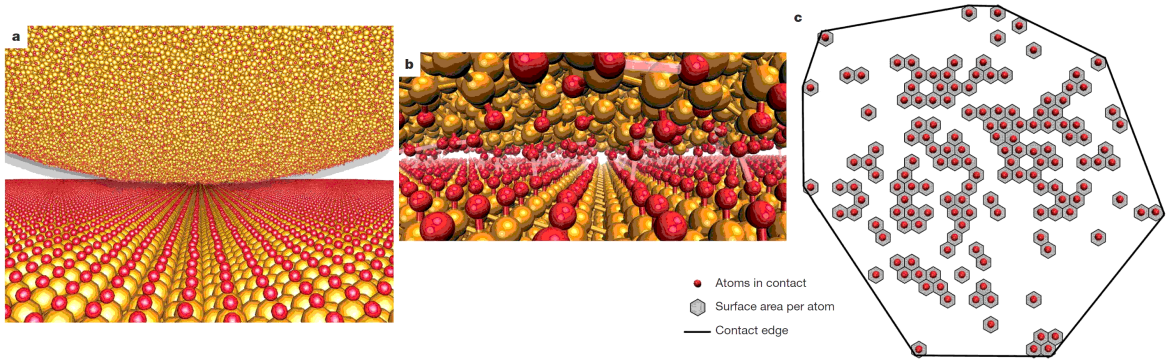}
\caption{\label{Slufi2009_Fig1}
  (a) MD simulated amorphous carbon tip over a diamond surface.
  Yellow and red atoms represent C and H, respectively.
  (b) A close view of the interface with highlighted covalent bonds (red
  and yellow sticks) and repulsive interactions (pink sticks)
  (c) An overall map of the contact area.
  Adapted from Ref.~\cite{Mo09}; Copyright (2009) by Nature Publishing
  Group.
}
\end{figure}

\subsubsection{AFM, nanotubes, and other systems.}
%ERIO:  QUESTA SOTTOSEZIONE SOTTO E' UN PO' MISERABILE...MA ANDARE IN DETTAGLIO DI TUTTI I CASI PORTEREBBE LONTANO....
%
Nowadays' computational capabilities even permit the atomistic simulation
of an entire AFM tip, enabling the understanding of several mechanisms
which are not describable by simplified PT-like models (see
Fig.~\ref{Slufi2009_Fig1}).
For example, it is possible to highlight the formation/rupture dynamics of
contacts in multi-asperity interfaces, and consequently estimate the true
contact area as a function of the apparent one.
Besides, it is possible to investigate the effect of the tip plasticity and
elasticity, which are of fundamental importance to define the
load-dependent contact area \cite{Mo09}, and as channels for dissipation
and wear \cite{Mulliah03, Klocke14, James14, Klemenz14, Liu15}.
This approach enables the bottom-up derivation of the linear scaling laws
of macroscopic friction with size, and their transition to the sublinear
ones for incommensurate nanosized contacts.
We can now understand that such transition takes place when the contact
roughness becomes large compared to the range of interfacial interactions
\cite{Mo09}.

In the study of repeated scratching of metallic surfaces by hard AFM tips,
widely employed in the field of micro/nano machining, MD simulations have
uncovered strongly non-linear trends of the frictional force with the feed
(i.e.\ the distance from the first groove), induced by lateral forces
exerting on the tip due to the substrate plasticity \cite{Yan07}.

It is also important to mention the simulations of nanotubes (NT), either
made of carbon or hexagonal BN, which, due to their extraordinary
mechanical and electronic properties, have been investigated with enormous
interest in the last decades.
Almost defectless NT can be formed nowadays with lengths of the order of
1~cm \cite{Zhang13}, and precise measurements of their mechanical and
frictional properties have started to appear in literature \cite{Garel12,
  Nigues14}.
Simulations of concentric nanotubes in relative motion (telescopic
sliding), have revealed the occurrence of well-defined velocities at which
friction is enhanced, corresponding to a washboard frequency resonating
with longitudinal \cite{Tangney06} or circular \cite{Zhang09} phonon modes,
leading to enhanced energy dissipation.
The frictional response becomes highly non-linear while approaching the
critical velocity and, contrary to macroscopic systems, washboard
resonances can arise at multiple velocities, especially for incommensurate
interfaces where more than one length scale may be in common to the
contacting surfaces \cite{Tangney06}.

The exceptional electro-mechanical properties of NTs have also been
investigated by various tip-based techniques, revealing a strong friction
anisotropy dictated by NTs orientation.
In this respect, simulation-assisted experiments of a sliding nanosized tip
over CNTs reveal that transversal friction is enhanced by a
hindered-rolling motion of the NT, with a consequent frictional dissipation
that is absent in the longitudinal sliding \cite{Lucas09}.
The same elastic deformation have been reportedly responsible of a reverse
stick-slip effect in the case of an AFM probe sliding over a super-lattice
CNT forest \cite{Lou08}.
Here, simulations reveal that the fast sticking is induced by the
penetration of the tip into the valley between the NTs and its interaction
with the tubes on both sides, causing an elastic shell buckling of the
CNTs.
In contrast, the gradual slipping occurs over much longer distance because
it includes both the sliding on the top of the NT and the energy release at
both sides of the graphitic wall.

\subsubsection{ Boundary lubricated sliding.}
\begin{figure}
\centerline{
\includegraphics[width=0.48\textwidth,clip=]{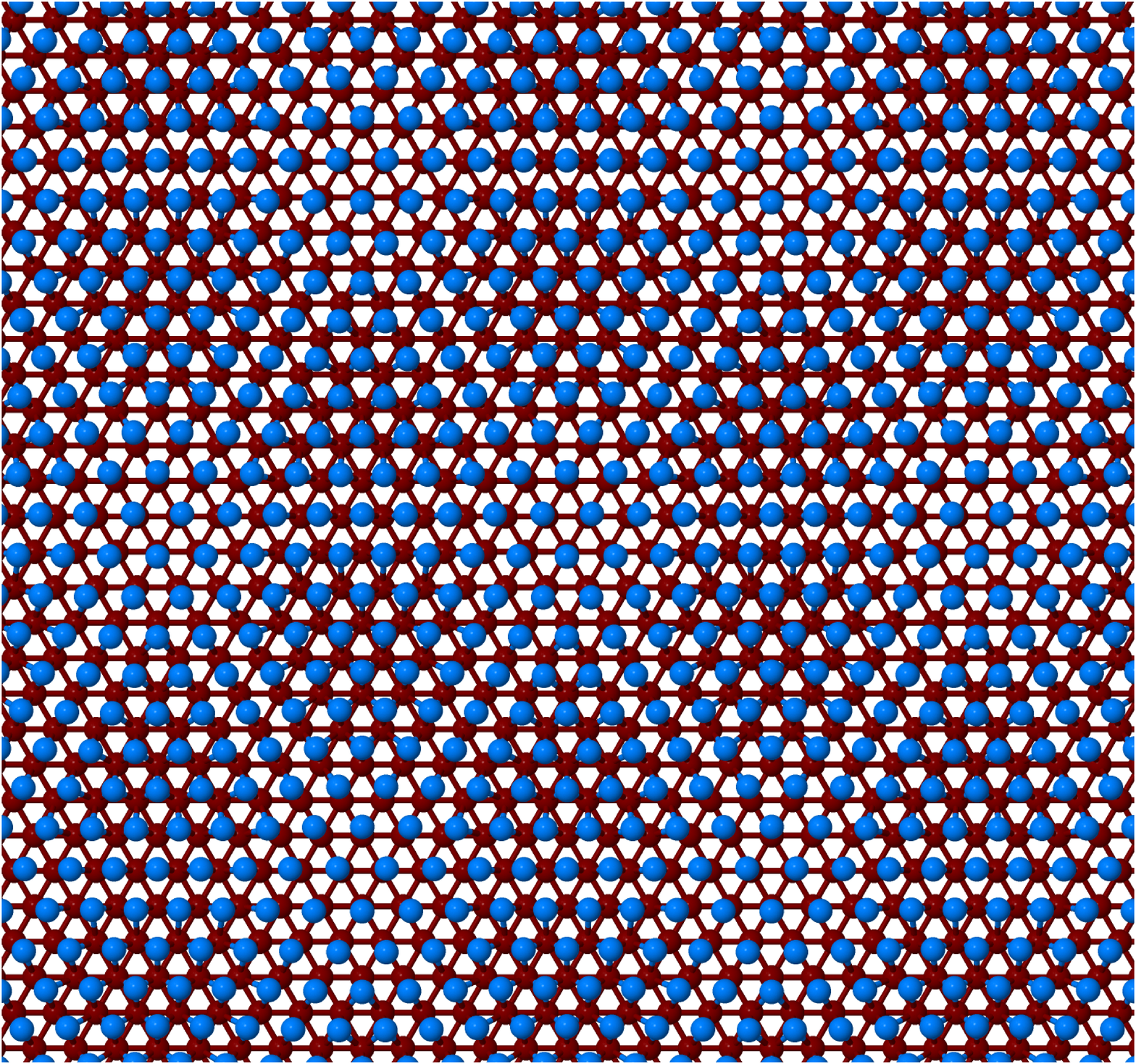}
}
\caption{\label{solid_lub}
  A snapshot of two layers of a 3D MD simulation, showing the hexagonal
  antisolitonic pattern formed at the boundary of a solid lubricant (light
  blue) in contact with lattice-mismatched crystalline surface (dark red).
  The lubricant is underdense, with a spacing misfit of $-11$\%.
  The Lennard-Jones interaction of this model favors the overlayer hollow
  sites: antisolitonic regions have locally unstable bridge- and top-site
  atoms.
  Other layers are omitted for clarity.
}
\end{figure}
%
%ERIO: NUOVA SOTTOSEZIONE. QUI ABBIAMO PIU' COSE DA DIRE....
%
When two sliding surfaces are separated by a thick lubricant film, as it
ordinarily happens under weak-load conditions, the tribological response of
the confined system is typically determined by the fluid viscosity.
In these cases of hydrodynamic lubrication, friction can be computed based
on the Navier-Stokes equations, which prescribe a monotonically increasing
kinetic friction as a function of the relative sliding speed
\cite{Szeri01}.
By contrast, at high load and low driving velocity, the lubricant may not
maintain a broad gap between the sliding surfaces, with the result that
solid-solid contact eventually occurs.
Prior to full squeezeout under pressure, as confirmed by experiments and
simulations, the intervening boundary film usually changes from liquid to
solid or nearly solid, exhibiting a layered structure prone to develop
finite static friction and a high-dissipative stick-slip dynamics in such
``boundary-lubrication'' regime.
Both SFA measurements \cite{Yosbizawa93} and MD investigations
\cite{Gao97a,Gao97b} have demonstrated sharp upward jumps of friction at
squeezeout transitions where the number of lubricant layers decreases from
$N$ to $N -1$.
Boundary-lubricated systems often display stick-slip dynamics during
tribological measurements associated with a significant value of the
friction dissipation.
As the load increases, however, it becomes harder and harder to squeeze out
an extra lubricant layer.
This hardening and increased difficulty of squeezeout reflects the
tendency to crystallization of the initially liquid lubricant and the
increased cost of the ``crater'' whose formation constitutes the nucleation
barrier of the transition \cite{Persson94a, Persson04a, Tartaglino06}.
Once it has happened, generally this $N \to N - 1$ “re-layering” transition
gives rise to an upward friction jump.
In principle however, upon re-layering of the solidified (structured)
confined film, the two-dimensional (2D) parallel crystalline-like order
could occasionally change under pressure toward a more favorable mismatched
(incommensurate) substrate-lubricant geometry.
In that case, as a result, sliding friction might actually switch downward
from highly-frictional stick-slip to smooth dynamical regimes
characteristic of incommensurate superlubric interfaces, with a highly
mobile 2D soliton pattern, of the type sketched in Fig.~\ref{solid_lub}.
So far this type of event has only been observed in simulations
\cite{Vanossi13}.

In SFA experiments, boundary-lubricated systems often display stick-slip
dynamics during tribological measurements, associated with a significant
value of the friction dissipation.
One cannot directly access the detailed film and interface rearrangements
giving rise to the stick-slip mesoscopic intermittent dynamics.
The mechanisms at play for the stick-slip dynamics in the
boundary-lubrication regime have been studied by MD investigations
\cite{Stevens93, Thompson90, Braun03, ZPBG2007}.
Several realistic models for lubrication layers were simulated
\cite{Braun06, Lorenz10a, Lorenz10b, Chandross04,Chandross08}.
The issue whether frictional shearing occurs through the middle of the
solid lubricant film, possibly accompanied by melting-freezing, or if it
forms a smooth shear band, or it occurs at the substrate-lubricant boundary,
is one which can be addressed by computer modeling.
Depending on the relative strength of the potentials governing the
lubricant-lubricant and lubricant-substrate interactions, a thin confined
film may exhibit a solid-like or liquid-like behavior under shear.
If the interaction with the substrates is weaker than the
lubricant-lubricant one, then sliding takes place mainly at the
surface-lubricant interfaces.
The lubricant film is then allowed to maintain or acquire a solid order.
If both the solid lubricant and the substrate are characterized by
nearly-perfect crystalline structures and these structures are mismatched
and/or misaligned \cite{ManiniBraun11, BraunManini11, deWijn11, deWijn12,
  vanWijk15}, then smooth superlubric sliding with reduced kinetic friction
ensues: in such conditions, solid lubrication can provide quite low
friction.
In practice however neither the substrates nor the lubricant are likely to
maintain undefected crystalline order.
Defects and/or impurities between the sliding surfaces, even if diluted to
a weak concentration, may suffice to induce pinning and finite static
friction \cite{Muser00}, thereby eliminating superlubricity.
In the opposite condition of prevailing lubricant-substrate interactions,
the surfaces are covered and protected from wear by lubricant monolayers:
sliding occurs inside the lubricant bulk.
In such condition the lubricant film can be led to melt during sliding;
alternatively, the layering imposed by the surfaces can remain solid, with
slips occurring in a layer-over-layer sliding \cite{Braun06,molshapeLin}.

\subsubsection{ Simulation of extreme frictional regimes.}
%
%ERIO: NUOVA SOTTOSEZIONE. ANCHE QUI ABBIAMO COSE...
%
Simulations are of particular value in the exploration of extreme
frictional regimes, that are difficult to access experimentally.
Among such extreme regimes, researchers have investigated or are
investigating high temperature, high speed, high pressure, and high plate
charging in ionic liquid lubrication.
Although for most of these conditions there still is no experimental
evidence to discuss, simulation has made some interesting predictions that
should become of future reference.

{\it High temperature.}
Close to the substrate melting point $T_{\rm m}$, its crystal surface may
or may not undergo surface melting -- the formation, in full thermal
equilibrium, of a thin liquid or quasi-liquid film at the substrate-vacuum
interface \cite{Tartaglino05}.
Either event influences importantly the contact of an AFM tip with the
surface.
Surface melting gives rise to a local jump-to-contact of the film with the
AFM tip, as was found both in experiments \cite{kuipers93} and in MD
simulations \cite{Tomagnini93}.
In that case, friction is expected to become hydrodynamic and
uninteresting.
More interesting is the case where the substrate surface does not undergo
surface melting, such as is the case for particularly well packed, stable
surfaces like Pb(111) or NaCl(100) \cite{Tartaglino05}.
For an AFM tip sliding on NaCl(100), frictional MD simulations suggested
two quite different outcomes depending on the frictional mode.
A sharp penetrating tip plowing the solid surface experiences a large
friction, which drops sharply when the substrate temperature is only
slightly below $T_{\rm m}$, so that the Joule heat suffices to raise
temperature locally and form a liquid drop accompanying and lubricating the
moving tip.
A blunt tip sliding wearlessly experiences instead a very small friction at
low temperature, counterintuitively surging and becoming large close to
$T_{\rm m}$, where the nonmelting surface lattice softens -- a phenomenon
analogous to that exhibited by flux lattices in type II superconductors
\cite{Zykova07}.

{\it High speed.}
Friction at high speed, of order of $1$~m/s or higher, is common in several
technologically relevant situations but is rarely addressed in nanoscale,
atomistically characterized situations, where velocity is more typically
$1~\mu$m/s, many orders of magnitude smaller.
As anticipated in Sec.~\ref{sizeissues:sec}, MD simulation is an ideal tool
for the study of friction in fast-sliding of nanosized systems.
Using gold clusters on graphite as test system, simulation has explored
high-speed friction, and especially differences and similarities from low
speed, examining the slowing down of a ballistically kicked cluster.
Both kinetic frictions are similarly viscous -- proportional to velocity.
However, they show just the opposite thermal dependence.
Whereas low speed (diffusive) friction decreases upon heating, when
diffusion increases, the high speed (ballistic) friction rises with
temperature, when thermal fluctuations of the contact increase
\cite{Guerra10}.

{\it High pressure.}
The local uniaxial pressure transmitted to a local contact by the overall
load on a slider may reach a hundred kbar, but is generally not very well
characterized, and the effects of pressure insufficiently explored.
MD simulation makes suggestions of different kinds.
First, pressure may provoke structural transformation of a crystalline
substrate (or slider) from its initial crystal structure to another.
As a recent simulation has shown \cite{Vanossi13} this will reflect in a
frictional jump, either up or down.
Second, pressure may bring a solid compound close enough to the chemical
stability limit for the frictional perturbation to cause bond breaking and
the beginning of chemical decomposition \cite{Crespo16}.
Third, pressure may lead to electronic or magnetic transformations, such as
insulator-metal transitions, and this may also in principle influence
friction.

{\it High plate charging in ionic-liquid lubrication.}
Ionic liquids are salts whose ions have such a large size that $T_{\rm m}$
falls below room temperature.
Experimental data have shown that friction across contacts lubricated by
ionic liquids depends on the state of electrical charging of the sliders
\cite{Sweeney12, Li14}.
MD simulations applied to heavily simplified ionic liquid models indicate
how this dependence can be ascribed to electrically induced structural
modifications at the slider-lubricant interface \cite{Fajardo15a,
  Capozza15a, Fajardo15b}.
For extreme plate charging, these modifications may even modify the
lubricant thickness, and also affect its whole molecular structure, with
strong predicted consequences on friction \cite{Capozza15b}.

%==================================================================
\section{Earthquakelike Models}
\label{multicontact:sec}

\begin{figure}
\centerline{
\includegraphics[width=0.5\textwidth,clip=]{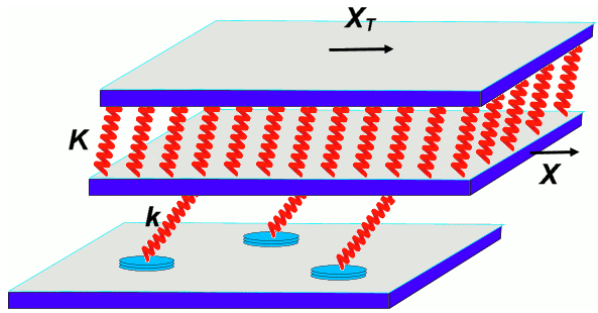}
}
\caption{\label{earthquakemodel}
A sketch of the earthquake-like model. Adapted from Ref.~\cite{Braun08},
Copyright (2008) by The American Physical Society.
}
\end{figure}

On meso- and macro-scales the interface between two bodies is quite
generally far from uniform.
When rough surfaces come into dry contact, the actual contacts occur at
asperities of different sizes, typically characterized by a fractal
distribution.
Even for a contact of ideally flat surfaces of polycrystal bodies,
different regions of the interface will be characterized by different local
values of the static friction due to structural or orientational domains
\cite{ManiniBraun11, BraunManini11}.
For a lubricated contact, different values of the local static friction may
appear due to patches of solidified lubricant or due to capillary bridges.
All these cases can be rationalized with the help of an earthquakelike (EQ)
model first introduced by Burridge and Knopoff \cite{Burridge67} to
describe real earthquakes.
The nature of two problems --- earthquakes and friction --- is very
similar: differences are restricted to their spatial-temporal scales:
kilometers and years to millenia in geology compared to nanometers and
seconds to hours in tribology.
The EQ model, also known as the spring-and-block model or the multi-contact
model, has been successfully used in many studies of friction
\cite{Barel10a, Barel10b, Braun02a, Persson95a, Filippov04, Braun09b,
  Braun09a, Braun11, Capozza12};
similar schemes have been used also to model the failure of fiber bundles
and faults \cite{Smalley85, Newman91, Newman01}.

In EQ models, two corrugated surfaces make contact only at a discrete set
of points, as shown schematically in Fig.~\ref{earthquakemodel}.
When the slider moves, a single point contact elongates elastically as a
spring, as long as the local shear force $f_i=k x_i$ ($x_i$ is the contact
stretching and $k$ is its elastic constant) remains below a threshold value
$f_{si}=k x_s$; then the contact breaks and slips for some distance, as
indeed was observed in tip-based microscopy experiments as well as in MD
simulations.
Then, either immediately or after some delay time, the contact is reformed
again with zero or lower stretching and a new threshold value.

The simplified version of the EQ model assumes that all contacts have the
same threshold $f_{si}$; such a model however corresponds to a singular
case and may lead to unphysical results \cite{Braun10}.
In a real situation, the contacts always have different thresholds with a
continuous distribution $P_c(x_s)$ of their static threshold elongations.
Therefore, when the upper block begins to advance, the forces acting
locally on each contact increase: at successive moments, the contacts begin
to snap in a sequence: weaker contacts break earlier, while the strongest
contact resists to the last.

EQ-like models are usually studied by simulation.
Nonetheless, the kinetics of the EQ model can be described by a master
equation (ME), occasionally known as Boltzmann equation or kinetic
equation.
In concrete, indicate the distribution of contact's stretchings $x$ by
$Q(x;X)$, when the sliding plate reaches position $X$ (see
Fig.~\ref{earthquakemodel}): the evolution of $Q(x;X)$ is described by the
equation \cite{Braun08, Braun10}
\begin{equation}
  \left[ \frac{\partial}{\partial X} +
  \frac{\partial}{\partial x}
  + P(x) \right] Q(x; X) = \delta (x) \,
  \int_{-\infty}^{\infty} d\xi \, P(\xi) \, Q(\xi; X) \,,
\label{int-eq02}
\end{equation}
where $P(x) \Delta X$ is the fraction of contacts which break at stretching
$x$ as a consequence of the plate advancing by $\Delta X$.
The ``rate'' $P(x)$ and the distribution of the breaking thresholds
$P_{c}(x)$ are connected by the relation
\begin{equation}
  P(x) = P_{c} (x) / J_c (x) \,,
\quad
  J_c (x) = \int_x^{\infty} d\xi \, P_{c} (\xi) \,,
\label{int-eq04}
\end{equation}
indicating that the fraction of contacts which snap when $X$ increases by
$\Delta X$ equals those that have their thresholds between $x$ and $x +
\Delta X$, divided by the total fraction $J_c(x)$ of contacts still
unbroken at stretching $x$.

The EQ model can be extended to account for thermal effects as well as the
ageing of contacts \cite{Braun10}; the latter requires an additional
equation to describe the increase of threshold values with the time of
stationary contact.
Analytic solutions of the ME are available and, in the smooth-sliding
regime, they provide us with the velocity and temperature dependence of the
kinetic friction force $F_k$ \cite{BP2011}.
Contrary to the Amontons--Coulomb laws, which state that (macroscopic)
friction is independent of velocity, the friction force in the EQ model
depends on the sliding speed.
At small driving velocity $F_k$ increases linearly with speed, $F_k (v)
\approx \gamma^* v$.
Indeed, if the slider moves slowly enough, thermal fluctuations will soon
or later break all the contacts.
The slower the slider the longer time any contact will be given to undergo
a fluctuation exceeding its respective threshold; therefore the smaller the
resulting kinetic friction force is.
This linear $F_k (v)$ dependence could be represented as a
(characteristically large) effective viscosity $\gamma^*$ of a ultrathin
lubricant film \cite{Braun09b, Braun11}.
At high driving velocities the friction force exhibits the opposite
behavior, it decreases when $v$ grows, $F_k (v) \sim f_s v^* /v$, due to
relaxation, which one can also call an ageing effect.
After snapping, a contact slips for a short time, then it stops and is
reformed again, growing in size and strength.
The faster the slider moves, the shorter time the contacts are left to be
reformed and to grow.
Overall therefore, the kinetic friction force $F_k(v)$ increases with $v$
at low $v$, up to a maximum at $v_0 \sim (f_s v^* /\gamma^*)^{1/2}$ and then
decreases.
At very high velocities $F_k(v)$ should eventually grow again due to an
increased damping in the slider bulk.
At intermediate speeds typical of experiments, the interplay of thermal and
ageing effects generates a weak (approximately logarithmic) $F_k (v)$
dependence, approximately consistent with the Amontons-Coulomb laws,
although the proper $F_k (v)$ dependence is rather difficult to detect in
experiments \cite{Barel10a, Barel10b}.

On the decreasing ("velocity-weakening") branch of the $F_k (v)$ dependence
at $v > v_0$ the slider motion may become unstable and change from smooth
sliding to stick-slip motion.
If the slider velocity increases due to a fluctuation, the friction force
decreases, and the slider accelerates.
This effect is studied usually with the help of phenomenological approach
(e.g., see \cite{Braun06} and references therein).
A detailed study of the EQ model \cite{Braun09b, Braun11, Braun10, Braun15}
shows that stick-slip for the multi-contact interface may appear, if and
only if two necessary conditions are satisfied.
First, the interface must exhibit an elastic instability.
When the slider moves, the contacts break but then are formed again, but
only if the reformed contacts build up a force sufficient to balance the
driving force, the motion will be stable.
Otherwise the slider will develop an elastic instability, and will keep
accelerating until the overall pulling spring force (of elasticity $K$, see
Fig.~\ref{earthquakemodel}) decreases enough to regain stability.
Second, the contacts must undergo ageing.
Once these conditions are satisfied, stick-slip will exists for an interval
of driving velocities only, $v_1 < v < v_2$, while for lower ($v < v_1$)
and higher ($v > v_2$) speeds the motion is smooth \cite{Braun15}.

The ME approach discussed above however assumed a rigid slider which is not
a proper model of a realistic extended system.
For a nonrigid slider, its elasticity produces contact-contact interaction:
as soon as a contact fails, the forces on nearby contacts must increase by
an amount $\delta \! f$.
This $\delta \! f (r)$ was shown \cite{Braun12a} to depend on the distance
$r$ from the failed contact as $\delta \! f (r) \propto r^{-1}$ at short
distance $r \ll \lambda$, and as $\delta \! f (r) \propto r^{-3}$ at long
distance.
The crossover length $\lambda \sim a^2 E/k$, known as the elastic
correlation length \cite{Caroli98, Braun12a}, depends on properties of both
the slider (its Young modulus $E$) and the interface (the mean separation
$a$ between nearby contacts and their average rigidity $k$).
Accordingly, a simpler model can be formulated that considers the slider as
rigid across regions of lateral size $\sim\lambda$, with the micro-contacts
inside each $\lambda$-sized area treated as a single effective macro-contact.
For this ``$\lambda$-contact'' the parameters can be evaluated by solving a
specific ME.

Numerics also showed, in addition, that a large fraction of the extra
inter-contact force concentrates behind and in front of the snapped
contact, implying that effectively the interface can be treated as a
one-dimensional chain of $\lambda$-contacts, at least approximately.
Now, if a $\lambda$-contact undergoes elastic instability, namely if at a
certain threshold stress the $\lambda$-contact snaps and then advances,
then the surrounding $\lambda$-contacts acquire an extra chance to also
fail: this mechanism results in a sequence of snaps propagating forward and
backward along the interface as in a domino effect.
The resulting dynamics of the chain of $\lambda$-contacts could then be
described by the Frenkel-Kontorova model of Sec.~\ref{FK:sec} above, but
replacing the sinusoidal substrate potential with a sawtooth profile,
i.e.\ a periodically-repeated array of inclined lines \cite{BP2012}.
This approach allows one to calculate the maximum and minimum shear stress
for the propagation of this self-healing crack (the minimum shear stress
coincides with the Griffith threshold \cite{Griffith21}), and also the
dependence of the crack velocity on the applied stress.
When a $\lambda$- sized contact fails at some point along the chain subject
to a uniform shear stress, two self-healing cracks leave the initial
snapping contact, propagating in opposite directions as divergent solitons
similar to the kink-antikink pair of Fig.~\ref{kinks}, until these cracks
either reach the boundary or meet another crack generated elsewhere.

On the other hand, nonuniform shear stress is relevant for experiments such
as those carried out by Fineberg's group \cite{Rubinstein04, Rubinstein06,
  Rubinstein12} with the slider pushed at its trailing edge: at this
location the shear stress is maximum and across the block it decreases with
the distance from the trailing edge.
In this system, the leftmost $\lambda$-contact is the most likely to fail
first, as the pushing force is increased.
This failure will result in an increased stress concentrating on the
successive $\lambda$-contact, which will fail as well.
This process will repeat itself until a self-consistent stress remains
below the breaking threshold everywhere in the slider.
As a result, the self-healing crack initiated at the trailing edge will
propagate along the interface for a certain distance $\Lambda$ (that can be
calculated \cite{BPS2014}), releasing the stress at its tail side, while
accumulating extra stress at its forward side.
When the pushing force further increases, a second crack starts at the
trailing edge, and can trigger a failure sequence in the pre-formed
stressed state, thus propagating to some extra distance.
These multiple cracks repeat themselves until they reach the slider leading
edge, resulting in a major collective slip.
Thus, at the sliding onset, several cracks advance along the interface,
with the whole slider undergoing multiple small forward slips (the
so-called precursors).
In experiment these precursors could be detected and could help predicting
the eventual large ``earthquake'' \cite{Braun09a,Rubinstein12}.

As mentioned at the beginning of this section, the EQ model was formulated
initially to explain earthquakes.
Actual earthquakes follow two approximate empirical laws:
the one named after Gutenberg-Richter \cite{Gutenberg54,Gutenberg56}
states that the number of earthquakes with the magnitude $\geq M$ scales
with the corresponding magnitudes according to a power law;
the Omori law \cite{Omori94} states that aftershocks occur with a frequency
decreasing roughly with the inverse time after the main shock.
EQ-like models discussed above can provide a rationale for both these laws
of seismology.
Specifically, the Gutenberg-Richter law can be understood as a direct
consequence of contact ageing \cite{BP2013}; the Omori law can be
interpreted in terms of cracks propagating to a finite distance --- after a
major earthquake, the stress is not released in full: a certain amount of
stress remains stored at a distance $\sim\Lambda$ from the main shock
\cite{BT2014}, where an aftershock is likely to occur some time later.

%==================================================================
\section{Conclusions}
\label{conclusions:sec}

The fascinating and multidisciplinary topic of microscopic friction, where
physics, engineering, chemistry and materials science meet to study the
process of converting mechanical energy irreversibly into heat, still lacks
fundamental understanding, and increasingly calls for well-designed
experiments and simulations carried out at well-characterized interfaces.
Although AFM, SFA, and QCM setups are providing insight in the high
nonlinear out-of-equilibrium interface processes at the small length
scales, these advanced experimental techniques still provide averaged
tribological data.
Overall physical quantities, such as the average static and kinetic
friction, the mean velocity and slip times, do not allow to tackle easily
the problem of relating the mesoscopic frictional response of a driven
system to the detailed microscopic dynamics and structural rearrangements
occurring at the confined interface under shear.

In this respect, by explicitly following and analyzing the dynamics of all
degrees of freedom at play in controlled numerical ``experiments'' with
interface geometry, sliding conditions, and interparticle interactions can
be tuned, mathematical modeling and computer simulations have proven
remarkably useful in the investigation of tribologic processes at the
atomic scale and are likely to extend their role in future frictional
studies.

Even though a large number of open questions remains to be addressed, these
modeling frameworks have provided effective insight in the nonlinear
microscopic mechanisms underlying the complex nature of static and kinetic
friction, the role of metastability, of crystalline incommensurability, and
of the interface geometry.
Each theoretical approach, from simplified descriptions, to extended
realistic MD and hybrid multiscale simulations, has limitations and
strengths, with specific abilities to address specific aspects of the
physical problem under consideration.
Thus, a robust prior understanding of the theoretical background is a basic
first step in deciding which modeling features deserve a specific
attention and which ones are rather irrelevant details, and then in
selecting the best methodological approach for a given problem.

Concluding, it is worth recalling that novel experimental approaches
address the intrinsic tribological difficulties of dealing with a buried
interface with a very limited control of the physical parameters of the
frictional system:
artificial systems consisting in optically trapped charged particles,
either cold ions in empty space or colloidal particles in a fluid solvent,
forced to slide over a laser-generated periodic potential profile.
Indeed, especially in 2D colloid sliding it is possible to follow each
particle in real time, like in MD simulations.
By knowing and, on top of that, tuning the properties of a sliding
interface, our physical understanding can expand significantly and open up
possibilities to control friction in nano- and micro-sized systems and
devices, with serious possibilities of bridging between nanoscale and
mesoscale sizes and phenomena.

%==================================================================
\section*{Acknowledgments}

Useful discussion and collaboration with
S.\ Zapperi
M.\ Urbakh,
J.\ Scheibert,
M.\ Peyrard,
B.\ Persson,
G.E.\ Santoro,
R.\ Capozza,
A.R.\ Bishop,
and
A.\ Benassi,
is gratefully acknowledged.
This work is partly funded by
the ERC Advanced Grant No. 320796-MODPHYSFRICT,
the Swiss National Science Foundation Sinergia CRSII2\_136287,
and by COST Action MP1303.
O.B.\ acknowledges partial support from the EGIDE/Dnipro grant No.\ 28225UH
and from the NASU ``RESURS'' program.

\section*{References}

%\bibliographystyle{unsrt}
%\bibliography{biblio}

\end{document}